\documentclass[fleqn,usenatbib]{mnras}

\usepackage{newtxtext,newtxmath}

\usepackage[T1]{fontenc}

\DeclareRobustCommand{\VAN}[3]{#2}
\let\VANthebibliography\thebibliography
\def\thebibliography{\DeclareRobustCommand{\VAN}[3]{##3}\VANthebibliography}

\usepackage{graphicx}	
\usepackage{subfigure}
\usepackage[flushleft]{threeparttable}

\newcommand{\1}{\,{\sc i}}
\newcommand{\2}{\,{\sc ii}}
\newcommand{\3}{\,{\sc iii}}
\newcommand{\4}{\,{\sc iv}}
\newcommand{\5}{\,{\sc v}}
\newcommand{\6}{\,{\sc vi}}

\newcommand{\sbs}{SBS 0335-052E}
\newcommand{\hst}{{\it HST}}
\newcommand{\al}{$\alpha$}

\newcommand{\tauV}{\mbox{$\hat{\tau}_V$}}
\newcommand{\msun}{\mbox{$M_\odot$}}

\newcommand{\Us}{\mbox{$U_\mathrm{S}$}}
\newcommand{\xid}{\mbox{$\xi_\mathrm{d}$}}
\newcommand{\nH}{\mbox{$n_{\mathrm{H}}$}}

\newcommand{\Hii}{\mbox{H\,{\sc ii}}}
\newcommand{\Zism}{\mbox{$Z_\mathrm{ISM}$}}

\newcommand{\beagle}{\textsc{beagle}}

\title[COS and MUSE observations of SBS 0335-052E]{Stars and gas in the most metal-poor galaxies I: COS and MUSE observations of SBS 0335-052E}

\author[A. Wofford et al.]
{\parbox{\textwidth}{Aida Wofford,$^{1}$\thanks{E-mail: awofford@astro.unam.mx}
Alba Vidal,$^{2,3}$
Anna Feltre,$^{4,5}$
Jacopo Chevallard,$^{2}$
St\'{e}phane Charlot,$^{2}$
Daniel P. Stark,$^{6}$
Edmund C. Herenz,$^{7}$
Matthew Hayes$^{8}$}
\\\\
\parbox{\textwidth}{$^{1}$Instituto de Astronom\'ia, Universidad Nacional Aut\'onoma de M\'exico, Unidad Acad\'emica en Ensenada, Km 103 Carr. Tijuana$-$Ensenada, Ensenada 22860, M\'exico\\
$^{2}$Sorbonne Universit\'es, UPMC-CNRS, UMR7095, Institut d'Astrophysique de Paris, F-75014 Paris, France\\
$^{3}$Laboratoire de Physique de l’Ecole normale supérieure, ENS, Université PSL, CNRS, Sorbonne Université, Université de Paris, F-75005 Paris, France\\
$^{4}$INAF – Osservatorio di Astrofisica e Scienza dello Spazio di Bologna, Via P. Gobetti 93/3, 40129 Bologna, Italy\\
$^{5}$Scuola Internazionale Superiore di Studi Avanzati (SISSA), Via Bonomea 265, I-34136, Trieste, Italy\\
$^{6}$Steward Observatory, University of Arizona, 933 N Cherry Ave, Tucson, AZ, 85721, USA\\
$^{7}$European Southern Observatory, Av. Alonso de C\'ordova 3107, 763 0355 Vitacura, Santiago, Chile\\
$^{8}$Stockholm University, Department of Astronomy and Oskar Klein Centre for Cosmoparticle Physics, AlbaNova University Centre, SE-10691, Stockholm}
}

\date{Accepted XXX. Received YYY; in original form ZZZ}

\pubyear{2015}

\begin{document}
\label{firstpage}
\pagerange{\pageref{firstpage}--\pageref{lastpage}}
\maketitle

\begin{abstract}
Among the nearest most metal-poor starburst-dwarf galaxies known, SBS 0335-052E is the most luminous in integrated nebular He\2~$\lambda$4686 emission. This makes it a unique target to test spectral synthesis models and spectral interpretation tools of the kind that will be used to interpret future rest-frame UV observations of primeval galaxies. Previous attempts to reproduce its He\2~$\lambda$4686 luminosity found that X-ray sources, shocks, and single Wolf-Rayet stars are not main contributors to the He\2-ionizing budget; and that only metal-free single rotating stars or binary stars with a top-heavy IMF and an unphysically-low metallicity can reproduce it. We present new UV (COS) and optical (MUSE) spectra which integrate the light of four super star clusters in \sbs. Nebular He\2, [C\3], C\3], C\4, and O\3]~UV emission lines with equivalent widths between 1.7 and 5~\AA, and a C\4\,$\lambda\lambda$1548, 1551 P-Cygni like profile are detected. Recent extremely-metal poor shock + precursor models and binary models fail to reproduce the observed optical emission-line ratios. We use different sets of UV and optical observables to test models of constant star formation with single non-rotating stars which account for very massive stars, as blueshifted O\5~$\lambda$1371 absorption is present.  Simultaneously fitting the fluxes of all high-ionization UV lines requires an unphysically-low metallicity. Fitting the P-Cygni like + nebular components of C\4\,$\lambda\lambda$1548, 1551 does not constrain the stellar metallicity and time since the beginning of star formation.  We obtain 12+log(O/H)$\,=7.45\pm0.04$ and log(C/O)$\,=-0.45^{+0.03}_{-0.04}$ for the galaxy. Model-testing would benefit from higher spatial resolution UV and optical spectroscopy of the galaxy. 
 
\end{abstract}



\section{Introduction}\label{sec:intro}

\textit{Extremely metal-poor massive ($\geq8\,M_\odot$) stars.} Stars with main-sequence (MS) masses above 20-25 $M_\odot$ are important because they dominate the ionizing flux of young stellar populations. Unfortunately, in the extremely-low metallicity regime, which is defined as $Z\leq$\,$Z_\odot/10$, where $Z$ is the fraction by mass of metals to all elements, the evolution of these stars and their chemical, radiative, and mechanical energy outputs are poorly understood. In this metallicity regime, we lack observations of individual stars, in particular far ultraviolet (FUV) spectroscopy, which enables the characterisation of stellar wind properties and the measurement of atmospheric abundances and ionizing luminosities \citep{Garcia2014}. One  looks for metal-deficient massive stars in H\,{\sc ii} regions with a low ionized-gas O/H ratio, where O/H is the fraction by number of oxygen to hydrogen atoms. This is because i) the atmospheres of MS massive stars have metal abundances which are similar to those of the H\,{\sc ii} regions in which they are embedded \citet{Asplund2009, Garcia-Rojas2014, Bouret2015}; ii) oxygen is the most abundant metal in nearby H\,{\sc ii} regions and optical oxygen lines are luminous in these regions; and iii) it is relatively easy to measure the O/H content of H\,{\sc ii} regions from the latter lines. Extremely metal-poor H\,{\sc ii} regions give their name to extremely metal-poor galaxies (XMPs), which are arbitrarily defined to have O/H$\,\leq$\,(O/H)$_\odot/10$ in the ionized gas \citep{Kunth2000}, or equivalently, using 12+log(O/H)$_\odot=8.69$ as the solar reference value \citep{Asplund2009}, 12+log(O/H)$\,\leq7.56$.\footnote{Hereafter, we we use the solar reference value of \cite{Asplund2009}, which corresponds to $Z_\odot=0.0134$, unless specified otherwise. We sometimes give abundances in parenthesis which correspond solar reference values of \cite{Caffau2011}, i.e., 12+log(O/H)$_\odot=8.83$ and $Z_\odot=0.0152$.} Unfortunately, the three known nearby XMPs where individual massive stars can be observed (Sextans A, SagDIG, and Leo P) are not strongly star-forming \citep{Camacho2016, Garcia2018, Garcia2019, Evans2019}. Thus, their stars do not significantly cover the parameter space of mass, luminosity, effective temperature, age, rotation, dynamical interaction in dense massive clusters, and in the case of stars in binary systems, orbital parameters and mass ratio combinations, which is relevant for either OB-type star astrophysics or as a reference to model distant starburst galaxies. The lack of empirical templates for extremely metal-poor massive stars severely limits our ability to understand progenitors of gravitational-wave events, long-Gamma Ray Bursts and Supernovae of types II/Ib/Ic; the integrated light of unresolved massive star populations; and the radiative, mechanical, and chemical feedback of massive star populations. In this context, rest-frame FUV observations of super star clusters (SSCs), also known as young massive clusters (YMCs), which are located in XMPs, provide critical constraints.

\textit{Extremely metal-poor SSCs.}  Super star clusters are dense aggregates of young ($\lesssim100$ Myr) stars with total stellar mass $\geq10^4\,M_\odot$, which are considered to be fundamental building blocks of galaxies \citep{Portegies2010}. In particular, the most massive SSCs are believed to be the progenitors of today's globular clusters \citep{Bastian2016}. Ultraviolet observations of SSCs in XMPs, reveal how extreme-low metallicity affects the integrated properties of massive star populations and their surrounding gas, and provide valuable constrains to spectral synthesis models, i.e., models of the integrated spectra of starlight which sometimes also include its reprocessing by interstellar gas and dust. Such models constitute the main tool for interpreting observations of unresolved stellar populations and galaxies, near and far.  

\textit{XMPs in the context of future large telescopes.} The James Webb Space Telescope (JWST) and Extremely Large Telescopes (ELTs), will obtain rest-frame UV spectra of thousands of galaxies in the re-ionization era, at redshifts between 6 and 15. Spectral synthesis models will be the main tool to interpret such observations, which will challenge our understanding of galaxies with extreme low metal content \citep{Byler2020}. Given that it only takes a few pair instability supernovae to reach a metallicity of 2\% by a redshift of $z\sim7$ \citep{Wise2012}, it is unlikely that we will ever be able to study distant, truly metal-free galaxies in detail. Thus, in order to understand how low metallicity affects the properties of stars and gas, it is essential to study SSCs in XMPs. Of course, XMPs are expected to also host older stellar populations which produced the oxygen that is currently present in the galaxy. This is the case in the prototypical XMP starburst, I Zw 18, which contains Red Giant Branch stars formed at least 1 Gyr ago \citep{Annibali2013}.

\textit{Low metallicity threshold of the nearby Universe.} Using SDSS DR7, \cite{Sanchez2016} found that XMPs have a low metallicity threshold of 12+log(O/H)=7.1. This threshold corresponds to 1/39th (or 1/54th) of the present-day photospheric oxygen abundance of the Sun. Oxygen abundance measurements in AGC 198691 \citep{Hirschauer2016} and SBS 0335-052W \citep{Izotov2009} push the threshold even lower to 12+log(O/H)=$7.0\pm0.03$, i.e., to 1/49th (or 1/68th) of the solar value.  Interestingly, the latter two XMPs have a metallicity which is similar to that of some Damped Lyman Alpha Systems at $z\sim5$ \citep{Rafelski2014}. The low metallicity content of XMPs remains to be explained. It is not clear if it is due to a low time-averaged star formation rate; the preferential escape of oxygen from their low gravitational potential well after type II SN explosions; accretion of metal-poor gas \citep{Filho2015}; or location in sparsely populated groups of galaxies where inflow of IGM gas is low and tidal interactions between galaxies are rare, leading to fewer episodes of star formation.  

\textit{Types of XMPs.} Galaxies which have been classified as XMPs comprise blue compact galaxies (BCGs), objects caught in a phase of substantial star-formation which appear blue and compact because their star-forming regions are large relative to the galaxy's projected size; blue compact dwarf (BCD) galaxies, which have similar properties to BCGs, but are fainter than $M_{\rm{B}}=-18$ mag \citep{Thuan1997}; dwarf irregular (dIs) galaxies, which can be considered extreme late-type spirals with large gaseous components \citep{Battaner2000}; and low surface brightness (LSBs) galaxies, which are diffuse galaxies defined as having surface brightnesses at least one order of magnitude lower than the dark night sky \citep{Impey1997}. These galaxies span the stellar mass range (log $M_\star/M_\odot=6-10$) and are still forming their stellar disk. Some authors consider as "true" XMPs objects with O/H values below what is expected from their absolute blue-band magnitudes or stellar masses \citep{Ekta2010, Pustilnik2016}. Here we adopt the broader definition of XMP. 

\textit{How to find XMPs?} Multiple methods have been used to identify XMPs. \cite{Morales-Luis2011} and \cite{Sanchez2016} searched for XMPs using spectra from the seventh data release of the Sloan Digital Sky Survey (SDSS DR7, \citealt{Abazajian2009}). They selected as candidate XMPs objects with weak [N\2] lines relative to H$\alpha$ or strong [O\3] $\lambda$4363 emission, and then measured the O/H of the objects. Strong nebular [O{\sc\,iii}] $\lambda$4363 emission indicates a high electron temperature of the emitting gas, which is attributed to inefficient cooling of the gas via metal emission lines. Based on this method, \citep{Sanchez2016} found that the fraction of XMPs relative to all emission line galaxies in SDSS DR7 is only 0.01\%. \cite{James2015} used a morphology-based approach to target faint blue systems with H\2 regions embedded in a diffuse continuum. Of the 12 candidate XMPs found via this approach, two were spectroscopically confirmed as XMPs. Some of the most metal-poor XMPs have been found by the Arecibo Legacy Fast ALFA (ALFALFA, \citealt{Giovanelli2005}) survey, which is a multi-wavelength study of galaxies with H\1 masses in the range of $10^6-10^7\,M_\odot$. More recently, \cite{Hsy2018} identified 94 candidate XMPs via photometric colours and morphologies in SDSS DR12 data, and confirmed 45 as XMPs via 3-m Lick Observatory and 10-m W.M. Keck Observatory optical spectroscopy. Using colour-colour diagrams, \cite{Senchyna2019} uncovered 53 candidate XMPs with uniformly high specific star formation rates in SDSS imaging at magnitudes 16<{\textit i'}<23, i.e. extending significantly below the completeness limits of the SDSS spectroscopic survey, and confirmed 32 as XMPs via Multiple Mirror Telescope spectroscopy. Finally, \cite{Kojima2019} have used a machine learning classifier in order find XMPs candidates in Subaru/HyperSuprime-Cam optical images that are about 100 times deeper than SDSS. Spectroscopic follow up of 10 candidates confirmed three galaxies as XMPs.

\textit{Why focusing on SBS 0335-052E?} Studies of the most metal-poor nearby XMPs in the UV face two challenges: the scarcity of suitable sources, and the difficulty of obtaining rest-frame UV spectra, as these require space-based observations. SBS 0335-052E is one of the nearest ($\sim$54.1 Mpc)\footnote{Luminosity distance from NED, based on the H$\alpha$ velocity from \cite{Moiseev2010})}, most metal-poor (largest oxygen abundance is 12+log(O/H)=7.338$\pm$0.012, \citealt{Izotov1999}) starburst XMP known. It is also one of  most well-studied. Although it is farther and more metal-rich  than the prototypical, I Zw 18, which is located at $\sim$18.2 Mpc \citep{Aloisi2007} and whose North-Western component has 12+log(O/H)=$7.17\pm0.04$ \citep{Skillman1993}, the integrated He\2 luminosity of \sbs~(1.87e39 erg\,s$^{-1}$, \citealt{Kehrig2018}) is larger than that of I Zw 18 (1.12e38 \,s$^{-1}$, \citealt{Kehrig2015}). This makes of \sbs~a unique target to test spectral synthesis models. We recently obtained high-quality UV spectra of SBS 0335-052E, as part of our pilot program aimed at testing the combined stellar population + photoionization models of \cite{Gutkin2016}. A major goal of our program is to understand the achievements and limitations of these models and various fitting techniques when applied to spatially unresolved observations of XMPs.

\textit{Outline of this paper.} In section~\ref{sec:target}, we summarize results from previous studies of \sbs, and explain the importance of this galaxy for extragalactic studies. In section~\ref{sec:observations}, we describe the UV and optical observations which are used in this work. In section~\ref{sec:ionization_source}, we provide evidence for massive stars being a plausible main source of ionizing photons in the observed region of the galaxy. In section~\ref{sec:models_tool_tests}, we describe the models and analysis tool. We also present results from comparing the models to the observations. In section~\ref{sec:discussion}, we discuss our findings. Finally, in section~\ref{sec:conclusion}, we summarize and conclude.

\section{SBS 0335-052E}\label{sec:target}

\textit{Previous observations}. Among the nearest and most metal-poor starburst XMPs known,  BCD galaxy SBS 0335-052E is one of the most well-studied along with I Zw 18, which held the record of the most metal-poor galaxy known for a long time \citep{Kunth2000}. SBS 0335-052E was discovered in the Second Byurakan Survey \citep{Izotov1990}. Previous observations of \sbs\ include, but are not limited to, the following wavelength ranges and telescopes/instruments: in the radio, Very Large Array (VLA, \citealt{Pustilnik2001, Johnson2009}), Giant Metrewave Radio Telescope \citep{Ekta2009}, and Atacama Large Millimeter Array (ALMA, \citealt{Hunt2014, Cormier2017});  in the far-infrared, \textit{Herschel} Pacs \citep{Remy-Ruyer2015}; in the mid-infrared, \textit{Spitzer} MIPS and IRS \citep{Wu2008}; in the near-infrared, \textit{Hubble Space Telescope (HST)} NICMOS \citep{Reines2008, Thompson2009}; in the optical, Multiple Mirror Telescope \citep{Izotov1997}, Keck II \citep{Izotov1999}, 3.6 m ESO \citep{Papaderos2006}, Special Astrophysical Observatory (SAO, \citealt{Moiseev2010}), Very Large Telescope (VLT) GIRAFFE \citep{Izotov2006}, VLT FORS1+UVES \citep{Izotov2009}, VLT Multi Unit Spectroscopic Explorer (MUSE) \citep{Herenz2017b}; in the ultraviolet, \textit{International Ultraviolet Explorer} \citep{Dufour1993}, \textit{Far Ultraviolet Spectroscopic Explorer} \citep{Grimes2009}, \textit{HST} FOS \citep{Garnett1995}, \textit{HST} GHRS \citep{Thuan1997}, \textit{HST} Cosmic Origins Spectrograph (COS \citealt{James2014}); in the  X-rays, \textit{Chandra} \citep{Thuan2004, Prestwich2013}.  

\textit{Environment.} SBS 0335-052E has a companion BCD, known as SBS 0335-052W, which is separated in the east-west direction by 84", i.e., by 22 kpc at the distance adopted by \cite{Pustilnik2001} ($\sim54$ Mpc). The galaxy resides not far from a rather large group, with only $\sim$50 kpc in projection from an $L_*$ spiral. The difference in radial velocities is within the group velocity dispersion. This group is at the border of a void \citep{Peebles2001}.

\textit{Ionized- and neutral-gas abundances and abundance ratios.} Oxygen abundance measurements in the ionized gas of SBS 0335-052E include those of \cite{Izotov1999, Papaderos2006, Izotov2006} and \cite{Izotov2009}. The latest measurement found variations over spatial scales of $\sim$1-2 kpc in the range 7.11 to 7.32 dex. Table~2 of \cite{Izotov2006} gives the relative abundances of N, Ne, S, Cl, Ar and Fe relative to O, in the ionized gas. The difference between ionized- and neutral-gas metal abundances in \sbs~is larger than for I Zw 18 (\citealt{Lebouteiller2013}; see also \citealt{James2014}). Finally, using the  O\,{\sc iii}] $\lambda$1666 and C\,{\sc iii}] $\lambda$1909 line fluxes measured with \textit{HST}'s Faint Object Spectrograph,\footnote{This instrument is no longer available on \textit{HST}.} \cite{Garnett1995} derive log(C/O$)=-0.94\pm0.17$. 

\textit{Dust and H\1 gas.} \cite{Thuan1997} report that dust is clearly present and spatially mixed with the SSCs in \sbs. According to \cite{Remy-Ruyer2015}, \sbs~has the lowest dust-to-stars mass ratio  of the Dwarf Galaxy Survey of \cite{Madden2013} (log([$M_{\rm dust}$/$M_{\rm stars}$=-4.53]). It has also been found that \sbs~has a high H{\sc\,i} to dust mass ratio for a BCD \citep{Dale2001}. 
\textit{Physical properties.} Table~1 of~\cite{Cormier2017} summarizes the main physical properties of the galaxy. \sbs~is a starburst galaxy with a total stellar mass of $5.6\times10^6\,M_\odot$ and a star formation rate of 0.7 $M_\odot$ yr$^{-1}$. It also has the highest specific star formation rate (log($sSFR$/yr)=-8.13) of the Dwarf Galaxy Survey of \cite{Madden2013} according to \cite{Remy-Ruyer2015}. At the distance of \sbs, the detection of a red giant branch, and thus an underlying old population, is not possible.

\begin{figure}\label{fig:sbs}

\centering\offinterlineskip
\includegraphics[width=0.49\columnwidth]{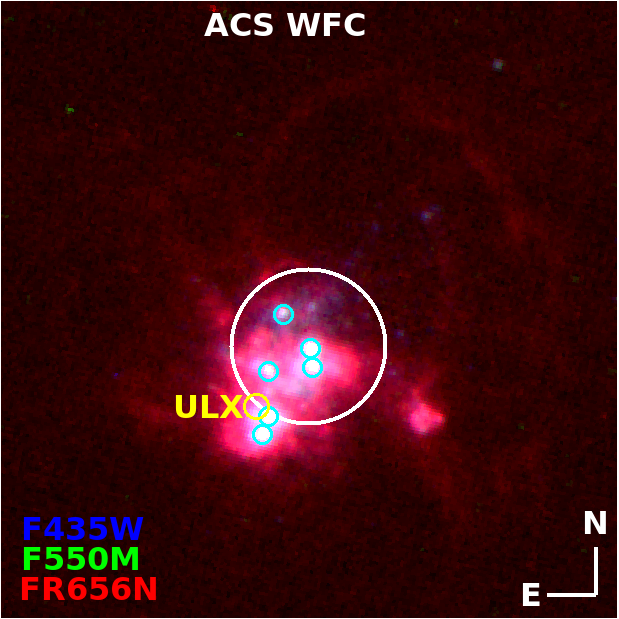}%
\includegraphics[width=0.49\columnwidth]{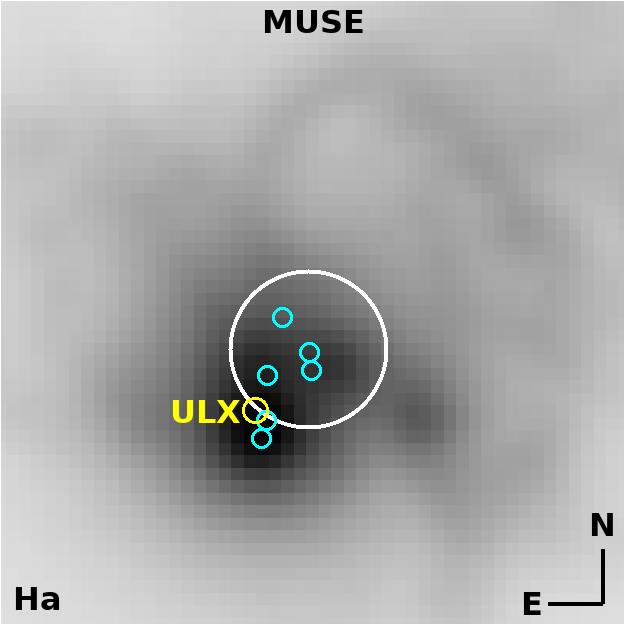}\\
\includegraphics[width=0.49\columnwidth]{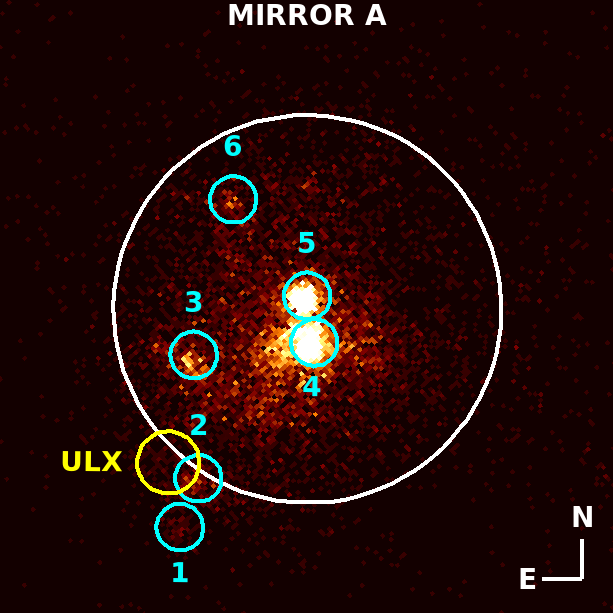}%
\includegraphics[width=0.49\columnwidth]{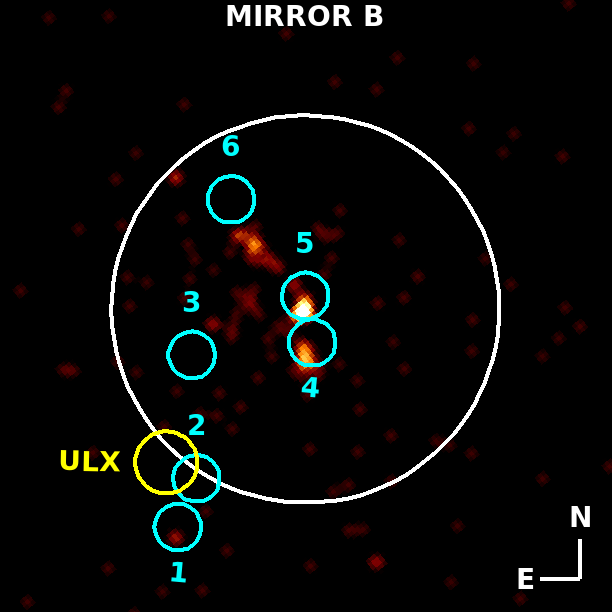}%
\caption{Top-left--. \textit{HST} ACS WFC RGB image of SBS 0335-052E composed from images of program 10575 (PI \"{O}estlin). The colour scale is logarithmic. In order to enhance the northern H$\alpha$ shell, the contrast and bias of the images were adjusted and the images were smoothed with a Gaussian kernel of three pixels in radius.  Top-right--. MUSE H$\alpha$ image obtained from the observed-frame wavelength range 6647 to 6655~\AA. The colour scale is logarithmic. Bottom-left--. \textit{HST} COS NUV Mirror A TA image (PID 11579, PI Aloisi). The colour scale is linear. Bottom-right--. COS NUV Mirror B TA image (PID 13788, PI Wofford). The colour scale is linear and the image is smoothed with a Gaussian kernel of three pixels in radius in order to enhance the signal from SSCs 4 and 5. In all panels: i) the largest circle (white) represents the COS PSA footprint and has a diameter of 2.5", i.e., 656 pc at the adopted distance to the galaxy; ii) the medium-sized circle (yellow) shows the location of the ULX which is reported in \citet{Prestwich2013} and has a diameter equal to \textit{Chandra}'s positional uncertainty, i.e., 0.4"; and iii) the smallest circles (cyan) indicate the positions of SSCs 1 to 6 which are identified in \citet{Thuan1997}.}
\end{figure}

\textit{Morphology.} \cite{Thuan1997} identified six compact SSCs and a supershell of radius $\sim$380 pc, which delineates a large supernova cavity. The left panel of Figure~\ref{fig:sbs} shows a false-colour RGB image of \sbs~where the supershell can be seen to the NE of the overlaid circle representing the COS aperture. Using an unsharp masking technique, \cite{Papaderos1998} discovered several lower-luminosity (and possibly more evolved) SSCs outside of the main conglomerate of the dominant six SSCs. Hereafter, we adopt the SSC IDs of \cite{Thuan1997} and references therein.
 
\textit{Importance for extra-galactic studies.} \cite{Izotov1999} observed a region of \sbs~which is centered on SSC 5 with the Keck II telescope. Using their emission line measurements, we find [O\,{\sc iii}] $\lambda$5007 / [O\,{\sc ii}] $\lambda$3727 = 15. Based on this ratio, \sbs~would be classified as a Green Pea (GP) galaxy \citep{Cardamone2009}.\footnote{Note that in the latter paper, the authors just call them Pea galaxies.} Green Pea galaxies are compact galaxies that are unresolved in SDSS images and were discovered in the citizen science project Galaxy Zoo.\footnote{http://zoo1.galaxyzoo.org/} The green colour of GPs is because [O\,{\sc iii}] $\lambda\lambda$4959, 5007 dominates the flux of the SDSS r-band, which is mapped to the green channel in the SDSS false-colour gri-band images. As shown in later papers, e.g. \cite{Izotov2011}, similar types of nearby galaxies exist also at redshifts where they do not appear green. Green pea like galaxies are the current best nearby analogs of high-z Lyman-alpha (Ly\al) emitters (LAEs), e.g. \cite{Yang2016, Yang2017, Jaskot2014, Henry2015, Izotov2016}. LAEs are an important population of star-forming galaxies at z > 2, increasing in fraction to constitute 60\% of Lyman break galaxies (LBGs) at redshifts z > 6 \citep{Stark12}, and they are used to probe reionization, e.g. \cite{Malhotra2004, Ouchi2010, Hu2010, Kashikawa11}. LAEs resemble some of the faint LBGs that dominate the luminosity density of the universe in the epoch of reionization \citep{Herenz2019} in size \citep{Malhotra2012}, dust, and stellar mass \citep{Finkelstein2010}. They have high line equivalent widths, e.g. \cite{Malhotra2002, Zheng2014}, indicating intense star formation despite their comparatively low stellar masses.\footnote{Note that there exist bright LBGs that show strong Ly\al as well as faint LBGs with no Ly\al \citep{Dunlop2013}.} In this context, understanding the origin of the high [O\,{\sc iii}]/[O\,{\sc ii}] ratio of \sbs~could shed light on some of the spectral properties of GP galaxies. It is thus important to understand the dominant ionizing source in SBS 0335-052E.

\textit{Ly\al~properties.} Interestingly, \cite{Thuan1997} and \cite{Kunth2003} report an absence of strong H\1 Ly\al~emission from \sbs, which is attributed to a combination of dust attenuation, redistribution of the Ly\al~photons by resonant scatterings over the whole area of the surrounding H\1~cloud, and the geometry of the cloud. Note that the column density of H\1~is very high, i.e., log[N(H\1) cm$^{-2}$]=20.15 \citep{James2014}. The UV spectrum of \hst~program 11579 (PI Aloisi) confirms the absence of significant Ly\al~emission from \sbs, as reported in \cite{James2014}.  The latter spectrum is reproduced in the top panel of Figure~\ref{fig:cos_spectrum}. In addition, extended diffuse Ly\al~is undetected around the galaxy
\citep{Ostlin2009}. \cite{Wofford2011} present Ly\al~line observations of 20 star-forming galaxies with a range of morphologies, oxygen abundances, and SFRs. All the galaxies in their sample with Ly\al~in emission have outflows of H\,{\sc i} gas. \cite{Rivera-Thorsen2015} present far-UV spectroscopy of 14 strongly star-forming star-forming galaxies at low redshift (0.028$<z<$0.18). They find that all galaxies in their sample with Ly\al~ escape show bulk outflow velocities of $\geq50$ km s$^{-1}$, although a number of their galaxies with similar velocities show little or no Ly\al~escape. \cite{James2014} find that in SBS 0335-052E the H\,{\sc i} gas is outflowing at $\sim30$ km s$^{-1}$ relative to the UV background in \sbs. In summary, \sbs~is a strongly star-forming galaxy with the oxygen line ratio of a GP galaxy, with outflowing H\,{\sc i} gas, but with no Ly\al~emission.  

\section{Observations}\label{sec:observations}

In this section, we present \hst~COS and {\it VLT} MUSE spectra of \sbs~which we use to i) discuss plausible ionizing sources for the interstellar gas in the galaxy, ii) test spectral synthesis models, and iii) derive physical properties of the stars and gas.  

\subsection{COS spectroscopy}\label{sub:cos}

As part of program 13788 (PI Wofford), we obtained integrated observations of \sbs's SSCs 3, 4, 5 and 6, centered on SSC 5, using \hst's COS medium resolution gratings, G160M and G185M ($\lambda/FWHM\sim16,000-20,000$), which cover the wavelength range from $1421$ to $1950$~\AA~(observed-frame), with some gaps due to unfilled space between the detector segments. The spectra were obtained with the circular Primary Science Aperture (PSA), which is 2.5" in diameter, and whose footprint on the galaxy is shown in the left panel of Figure~\ref{fig:sbs}. The latter figure also shows the locations of the six SSCs identified by  \cite{Thuan1997}, and the ULX which is reported in \cite{Prestwich2013}. Our data extend the wavelength coverage of \hst~program 11579 (PI Aloisi), which uses COS G130M, also centered on SSC 5, to cover the  wavelength range $\sim1130-1420$ \AA~(observed-frame). \cite{James2014} use UV absorption lines in the G130M spectrum to derive metal abundances in the neutral gas. In this work, we use the G130M spectrum to look for signatures of Very Massive Stars (VMSs), which are stars with masses above 100$\,M_\odot$ \citep{Vink2012}. We also use the G130M spectrum to obtain a better fit to the stellar continuum and check that the flux levels of all datasets match, including the optical spectrum.

We used the following observing sequence. 1) Perform target-acquisition (TA) procedures in order to ensure that the target is well centered in the PSA. This avoids throughput losses, produces reliable wavelength zero points, and outputs an image of the target over an NUV band of wavelength range, $1700-3200$ \AA~(observed-frame). In order to preserve the NUV detector, the TA image was obtained with MIRRORB, which offers an attenuation factor of $\sim25$ relative to MIRRORA. For reference, the plate scale of the NUV detector is 0.075 arcsec. 2) Obtain spectra in TIME-TAG mode using the FP-POS=ALL setting and FLASH=YES so that the spectra can be corrected for flat-field anomalies and OSM drifts. 3) Obtain additional spectra during subsequent orbits in order to achieve the desired signal-to-noise ratio. Note that for the G130M observations, MIRRORA was used for the TA. The TA images of programs 11579 and 13788  are shown in the middle and right panels of Figure~\ref{fig:sbs} with the PSA footprint overlaid and the location of the SSCs identified by the IDs of \cite{Thuan1997}. In spite of the lower SNR of our MIRRORB image, clusters 4 and 5 are clearly detected. This is expected as they are the brightest in the UV, which can also be seen in figure 1 of \cite{Kunth2003}, which shows \hst~ACS F140LP and F122M UV images. According to table 2 of \cite{Thuan1997}, SSCs 1 and 2, which are located just outside of the COS aperture, are the brightest in the optical. 

We retrieved the individual G160M and G185M spectral data sets from the Mikulski Archive for Space Telescopes and initially processed them on-the-fly with version 3.2.1 (2017-04-28) of the CalCOS pipeline. We also retrieved the G130M data set, which was processed through Calcos version 3.0 (2014-10-30). Table~\ref{tab:obs_log} provides a summary of the observations which are presented in Figure~\ref{fig:sbs} and this work.  CalCOS processes the data for detector noise, thermal drifts, geometric distortions, orbital Doppler shifts, count-rate nonlinearity, and pixel-to-pixel variations in sensitivity. A standard wavelength scale is applied using the on board wavelength calibration. The G160M data fall on two detector segments, A and B. The two FUV segments are processed independently. The G185M data fall on three detector segments, A, B, and C. For each grating, multiple exposures with the same grating and central wavelength are contained within a visit, these are combined into a single summed spectrum. The final products are one-dimensional, flux calibrated, heliocentric-velocity corrected, background subtracted, and combined spectra which are stored in x1dsum files. We worked directly with the x1dsum files. By fitting a model for the stellar continuum of a Simple Stellar Population (SSP) to the G130M + G160M + G185M data, we noted a mismatch by a factor of 1.27 between the continuum levels of the G160M and G185M observations at 1715 \AA, in the wavelength region of overlap between these gratings. The Space Telescope Science institute informed us that a likely cause for this mismatch is that the COS calibration files are overestimating the loss in sensitivity in the NUV gratings, which is why the flux in the NUV G185M data is slightly lower than in the G160M data. The COS team is working on this, and hopefully will release new reference files soon. For this reason, we multiplied the G185M flux array by 1.27.

For a point-source well-centered on the PSA, the instrumental $FWHM$ values of the G160M and G185M gratings are 0.09 \AA~($\sim$18 km\,s$^{-1}$ at 1500 \AA) and 0.11 \AA~($\sim$22 km\,s$^{-1}$ at 1500 \AA), respectively, as derived by multiplying the dispersions of the gratings (0.012 \AA~pix$^{-1}$ and 0.037 \AA~pix$^{-1}$, respectively) by the sizes of the spectral resolution elements of their associated detectors (6 pixels for FUV XDL and 3 pixels for NUV MAMA). Since CalCOS outputs oversampled data, we resample the data to the nominal dispersions of the gratings. Wavelength zero point shifts can occur due to imperfect centering and the flux profiles of the galaxies in the PSA. For the data taken with G160M, it is possible to check for wavelength offsets by determining the centroids of Galactic UV absorption lines towards our targets; with the caveat that negative offsets of $\le-50$ km s$^{-1}$ are likely due to intermediate or high-velocity clouds (IVC or HVC, respectively; e.g., \citealt{Wofford2011}). To check for offsets, we use the Si\,{\sc ii} $\lambda$1527 and Al\,{\sc ii} $\lambda$1671 Milky Way (MW) absorption lines. We find velocity blueshifts of -20 and -25 km s$^{-1}$, respectively. We do not correct the observations for these negative offsets which might be explained by Milky Way (MW) infalling clouds. We use the $FWHM$ of the above two Galactic UV absorptions as the effective spectral resolution of the observations. We find $FWHM$ values of 0.54~\AA ($\sim$106 km\,s$^{-1}$) and $\sim$0.49~\AA ($\sim$89 km\,s$^{-1}$) for Si\,{\sc ii} $\lambda$1527 and Al\,{\sc ii} $\lambda$1671, respectively, i.e., an average of $FWHM$ value of 0.51 \AA~($\sim$97 km\,s$^{-1}$ at 1600 \AA). We use the {\tt IDL} routine gauss\_smooth.pro to smooth the data in order to match the effective spectral resolution given by the MW lines. 

\begin{figure*}\label{fig:cos_spectrum}

\begin{subfigure}
  \centering
        \includegraphics[width=2\columnwidth]{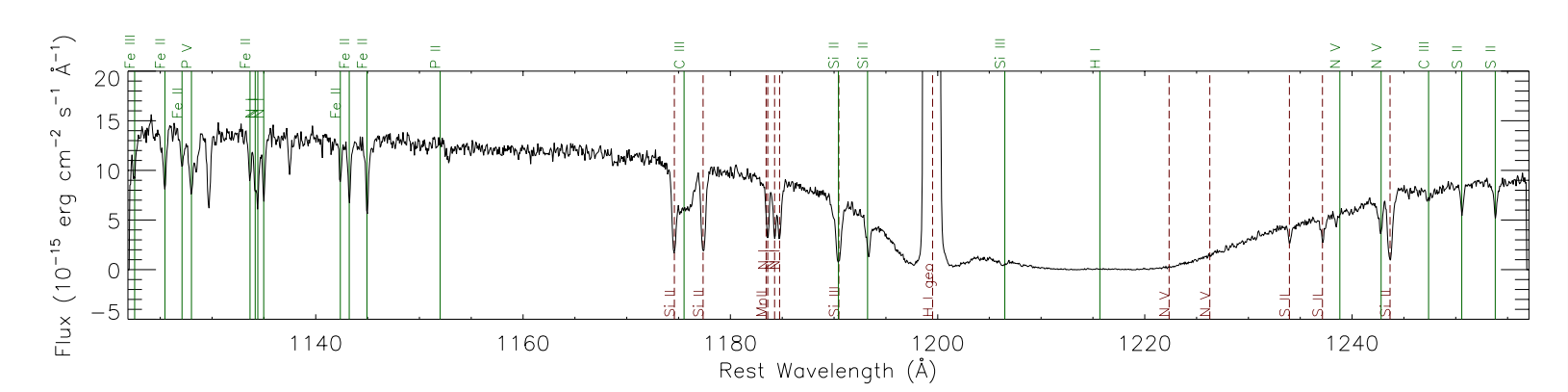}  
\end{subfigure}
\begin{subfigure}
  \centering
    \includegraphics[width=2\columnwidth]{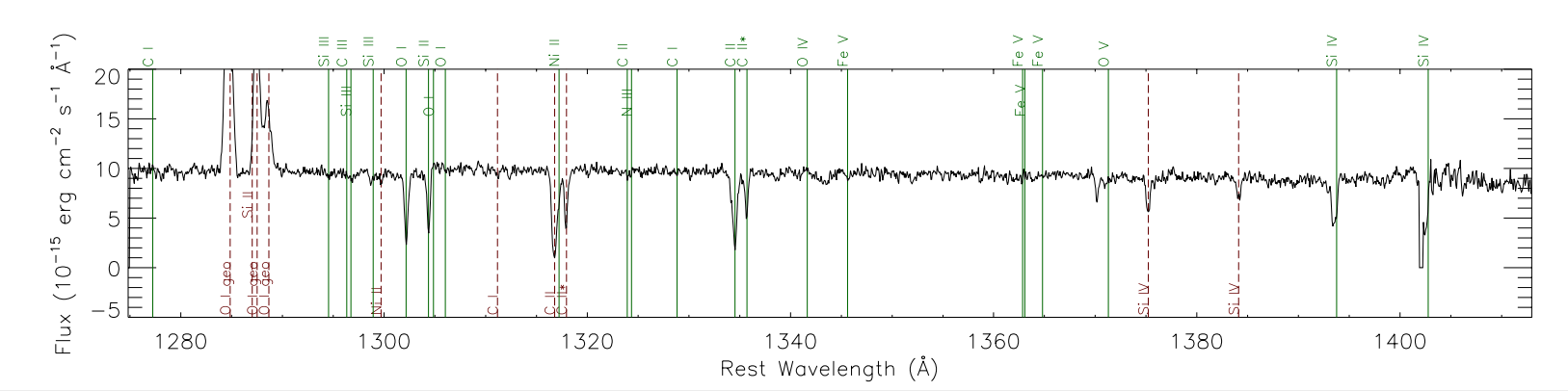}
\end{subfigure}
\begin{subfigure}
  \centering
    \includegraphics[width=2\columnwidth]{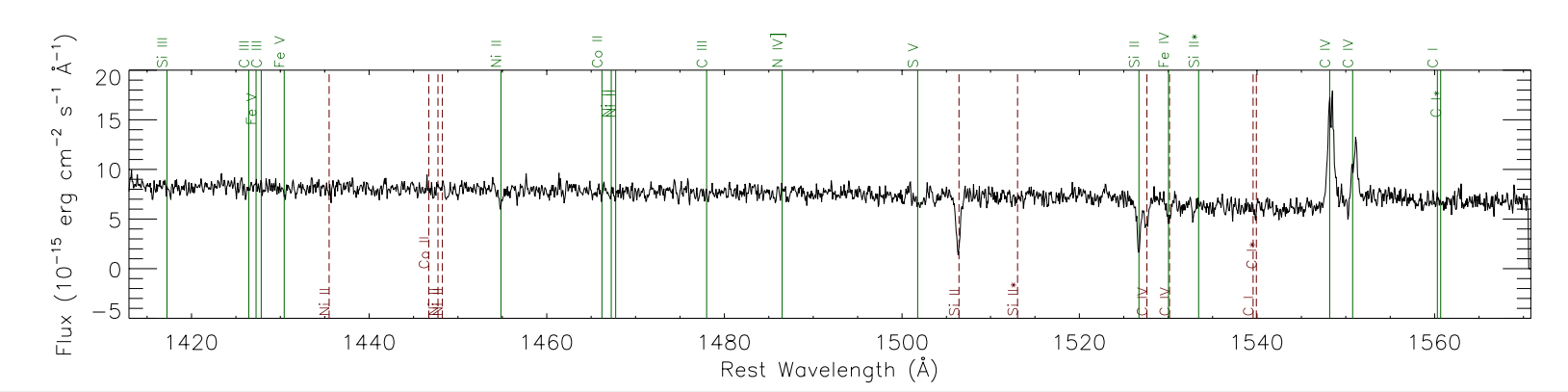}
\end{subfigure}
\begin{subfigure}
  \centering
\includegraphics[width=2\columnwidth]{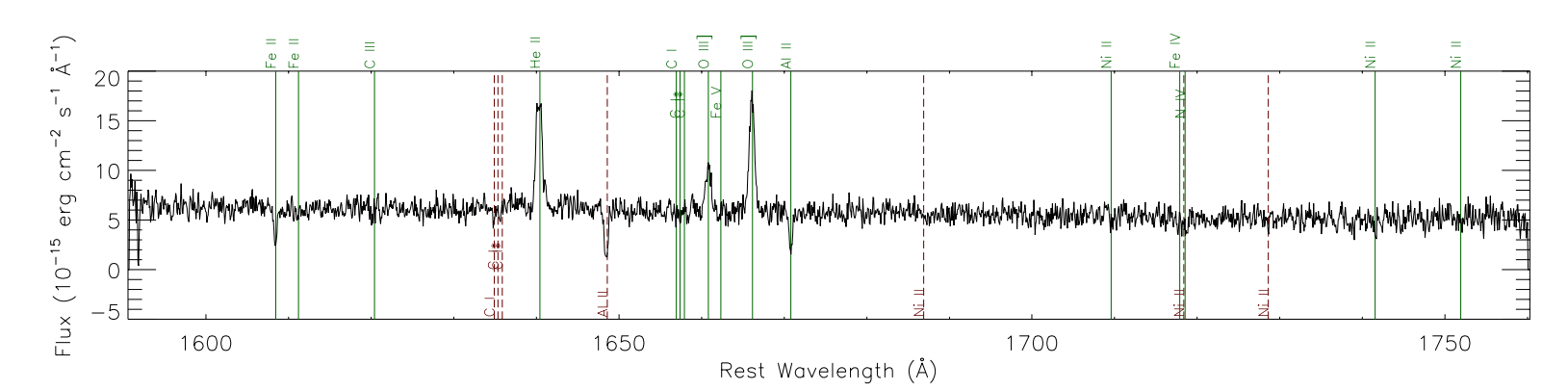}
\end{subfigure}
\begin{subfigure}
  \centering
    \includegraphics[width=2\columnwidth]{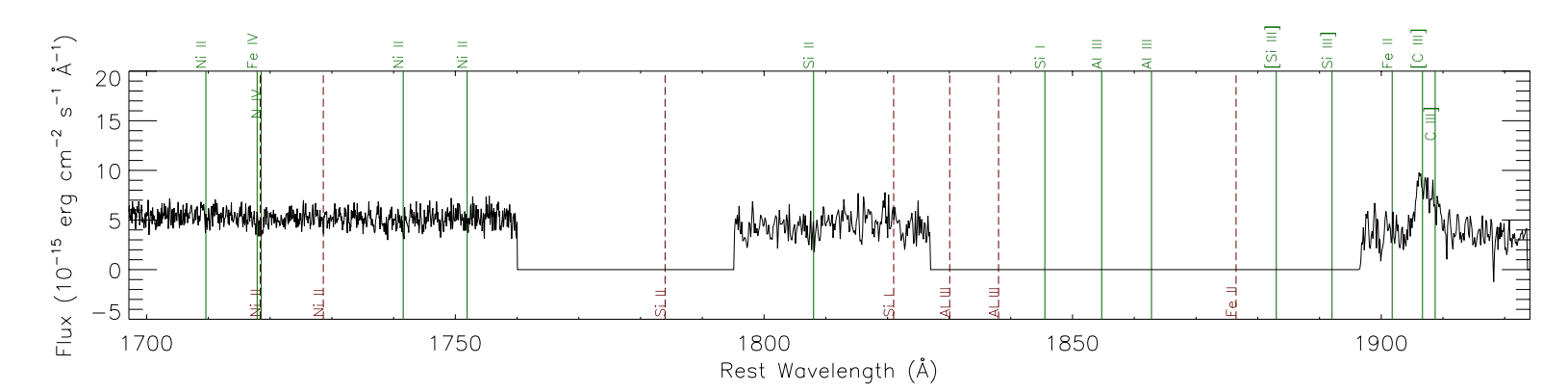}
\end{subfigure}
\caption{Archival G130M (PI Aloisi, PID 11579) and new G160M + G185M (PI Wofford, PID 13788) COS spectroscopy of \sbs. From top to bottom we show G130M/seg B, G130M/seg A, G160M/seg B, G160M/seg A, and G185M/seg A+B+C. The spectra are corrected for foreground reddening and redshift, and smoothed to the effective spectral resolution of the G160M data, i.e., 0.51~\AA. Intrinsic spectral features are marked with solid-green lines and labeled at the top of each panel. Foreground features are marked with dashed-brown lines and labeled at the bottom of each panel. We use the line list of \citet{James2014} below $1160\,$~\AA and that of \citet{Leitherer2011} above $1160\,$\AA. Note the detection of blueshifted O\5~$\lambda$1371 (plausibly from VMSs), photospheric S\5~$\lambda$1502 from massive stars, and nebular high-ionization UV emission lines.}
\end{figure*}

Figure~\ref{fig:cos_spectrum} shows the \sbs~COS G130M + G160M + G185M spectra from programs 11579 and 13788. When looking at the G130M spectra, one can see the three contributions to the Ly\al\,$\lambda1216$ profile: the MW Ly\al~absorption, the Earth's geocoronal Ly\al~emission, and the Ly\al~absorption which is intrinsic to \sbs~and is reported in \cite{James2014}. The G130M data also show blueshifted O\5\,$\lambda$1371 absorption, which has been attributed to the presence of VMSs in three metal-poor galaxies \citep{Crowther2016, Wofford2014, Smith2016}. The spectral morphology of O\5\,$\lambda$1371 is different than in the above galaxies owing to the much lower metallicity of SBS 0335-052E. Since O\5\,$\lambda$1371 is observed in early main sequence \citep{Bouret2013} and early O If stars \citep{Walborn1985, Walborn1995} as well as VMS, the association of blueshifted O\5\,$\lambda$1371 with VMS in \sbs~is not definitive. By looking at the G160M and G185M data, one can see that the high-ionization UV emission lines C\4\,$\lambda\lambda$1549, 1551,  He\2\,$\lambda$1640, O\3]\, $\lambda\lambda$1661, 1666, and [C\3], C\3]\, $\lambda\lambda$1907, 1909 are clearly detected. Previous lower spectral resolution observations of the C\,{\sc iv} and O\,{\sc iii}] doublets in \sbs~are shown in figure 1 of \cite{Dufour1993} and figure 1 of \cite{Garnett1995}, respectively. The first observation (PI Terlevich, PID NE077), was obtained with the SWP camera on board the {{\it International Ultraviolet Explorer}} using a 10"x20" oval aperture. It covers the wavelength range from $\sim$1150 to $\sim$2000 \AA~(observed frame). The second observation was obtained with \hst's Faint Object Spectrograph using a 1" circular aperture (PI Skillman, PID 3840). It covers the wavelength range from $\sim$1600 to $\sim$2000 \AA~(observed frame). With our G160M observation (0.51~\AA~at 1600 \AA), we resolve for the first time the C\,{\sc iv} and O\,{\sc iii}] doublets. Finally, we detect S\5$\,\lambda$1502 absorption from the photospheres of massive O and B stars \citep{Walborn1985, Walborn1995, deMello2000} at the 2$\sigma$ level. 

\begin{table*}\label{tab:obs_log}

\centering
\caption{Observation log of \textit{HST} observations used in this work. Columns. (1) ID of dataset. (2) J2000 right ascension in: hours minutes seconds. (3) J2000 declination in format: degrees minutes seconds. (4) Start time of observations in format: yy-mm-dd hh:mm:ss.00. (5) Exposure time. (6) Configuration. Wavelength ranges in \AA~of mirrors and gratings: MIRRA \& MIRRB, $1650-3200$; G130M segments, B[$1137-1274$], A[$1292-1432$]; G160M segments, B[$1421-1592$], A[$1612-1784$]; G185M segments, A[$1720-1753$], B[$1819-1852$], C[$1917-1950$].(7) Central wavelength of filter or grating. (8) Program ID and name of PI.}
\begin{tabular}{llllllll} 
\hline\hline
Dataset & RA & Dec & Start Time & Exp Time & Instrument/Aperture/ & $\lambda_{\rm{cen}}$ & PID/PI \\
\hfill & J2000 & J2000 & y:m:d h:m:s & s  & Filter or Grating & \AA & \hfill \\
\hline
J9FVA3030 & 03 37 44.000 & -05 02 40.00 & 2006-08-22 01:14:31 & 680 & ACS/WFC1/FR656N & 6651 & 10575/Oestlin \\
J9FVA3010 & 03 37 44.000 & -05 02 40.00 & 2006-08-22 00:44:12 & 430 & ACS/WFC1/F550M & 5581 & 10575/Oestlin \\
J9FVA3020 & 03 37 44.003 & -05 02 39.86 & 2006-08-22 00:57:14 & 680 & ACS/WFC1/F435W & 4330 & 10575/Oestlin \\
LB7H91KEQ & 03 37 43.980 & -05 02 38.90 & 2010-03-02 05:11:27 & 40 & COS/PSA/MIRRA & NUV & 11579/Aloisi \\
LB7H91010 & 03 37 43.980 & -05 02 38.90 & 2010-03-02 05:17:17 & 9534 & COS/PSA/G130M & 1291 & 11579/Aloisi\\ 
LCNE03SHQ & 03 37 43.980 & -05 02 38.90 & 2015-03-01 06:22:28 & 3 & COS/PSA/MIRRB & NUV & 13788/Wofford\\
LCNE03010 & 03 37 43.980 & -05 02 38.90 & 2015-03-01 06:28:31 & 4956 & COS/PSA/G160M & 1611 & 13788/Wofford\\ 
LB7H91010 & 03 37 43.980 & -05 02 38.90 & 2015-03-01 09:26:12 & 5675 & COS/PSA/G185M & 1835 & 13788/Wofford\\
LDN709020 & 03 37 43.980 & -05 02 38.90 & 2018-08-29 18:52:07 & 5605 & COS/PSA/G160M & 1623 & 15193/Aloisi\\
LDN759020 & 03 37 43.980 & -05 02 38.90 & 2019-02-14 06:49:25 & 5345 & COS/PSA/G160M & 1623 & 15193/Aloisi\\
\hline
\end{tabular}
\end{table*}

\subsection{MUSE spectroscopy}\label{sub:muse_obs}

SBS 0335-052E was observed with VLT/MUSE \citep{Bacon2010} in service
mode on November 16 and 17, 2015.  Sky conditions were clear to
photometric and the DIMM seeing FWHM varied between 0.9 and 1.2
arcsec.  The total open shutter time on target was 5680\,s, split into $8\times710$\,s individual exposures.  As recommended in the MUSE users manual, the spectrograph's field of view was rotated by 90$^\circ$ and small dither offsets were applied for each exposure.  For flux calibration the white dwarf GD\,71 \citep{Bohlin1995} was observed on November 17. On the same date, a twilight flat field was taken. Each open-shutter exposure is associated to a set of calibration exposures (bias frames, lamp flats, illumination flat, and arc-lamp frames) taken closest in time.

Data reduction of the MUSE observations was performed with the MUSE
data reduction system \citep[DRS][]{Weilbacher2014}, version 1.6.2.
We first used the calibration tasks {\tt{muse\_bias}},
{\tt{muse\_flat}}, {\tt{muse\_wavecal}}, {\tt{muse\_lsf}}, and
{\tt{muse\_twilight}} to create master bias frames, trace tables and
master flats, wavelength calibration tables, line spread function
images, and a twilight flat datacube, respectively.  With those
calibration data products, we then reduced all open-shutter exposures
into so-called pixtables with the DRS task {\tt{muse\_scibasic}}.
Since the instrument illumination has been shown to be temperature dependent, we ensured that we always applied the illumination flats taken
closest in time to the actual observation.  The standard
star pixtable was then fed into the task {\tt{muse\_standard}} to
generate a response curve for absolute flux calibration.  Thereafter,
we fed the response curve and individual science exposure pixtables
into the task {\tt{muse\_scipost}}, which calibrates the flux,
subtracts the sky spectrum, and astrometrically calibrates the
pixtables.  Lastly, using the DRS tasks {\tt{muse\_exp\_align}}
{\tt{muse\_exp\_combine}}, the fully calibrated pixtables were
combined and resampled onto a common $317 \times 317 \times 3800$ grid
(a so-called datacube), where the first two axis represent a spatial
position on the sky (regularly sampled at 0.2 arcsec $\times0.2$ arcsec),
and the third axis represents the spectral axis from 4600\AA{} to
9160\AA{} linearly sampled with 1.2~\AA steps.

The MUSE datacube of SBS\,0335-052E was already exploited by
\cite{Herenz2017b} to discover large extended H$\alpha$ and [OIII]
filaments in the halo of this galaxy. Owing to the unexpected large extend of nebular emission around this galaxy, the MUSE DRS sky-subtraction algorithm initially oversubtracted H$\alpha$ and [OIII] emission.  As described in \cite{Herenz2017b}, to correct for this, the resulting output sky-spectra from the DRS were first modified by interpolation over the affected wavelengths and subsequently fed into a second iteration of {\tt{muse\_scipost}}, as input
sky-spectra. The uncorrected MUSE datacube was exploited by \cite{Kehrig2018} to show that the total He\2 flux of the galaxy, including regions outside of our aperture, can only be produced by either single, rotating metal-free stars or a binary population with Z$\sim$10$^{-5}$ and a "top-heavy" IMF. 

For the present study, we use the corrected datacube to extract a spectrum at
the position of {\it{HST}}/COS. Therefore, we summed all spectral pixels of
the MUSE datacube within the COS PSA aperture.  We also utilise a continuum subtracted datacube to pinpoint spatially the faint Wolf-Rayet (WR) emission features (sec.~\ref{sub:wr}). Therefore we subtract an in spectral direction median-filtered (filter width = 180 \AA{}) continuum-only cube \citep[see Sect.~4.1 in][]{Herenz2017a}. 

\begin{figure*}\label{fig:muse_spectrum}
 
 \begin{subfigure}
  \centering
\includegraphics[width=2\columnwidth]{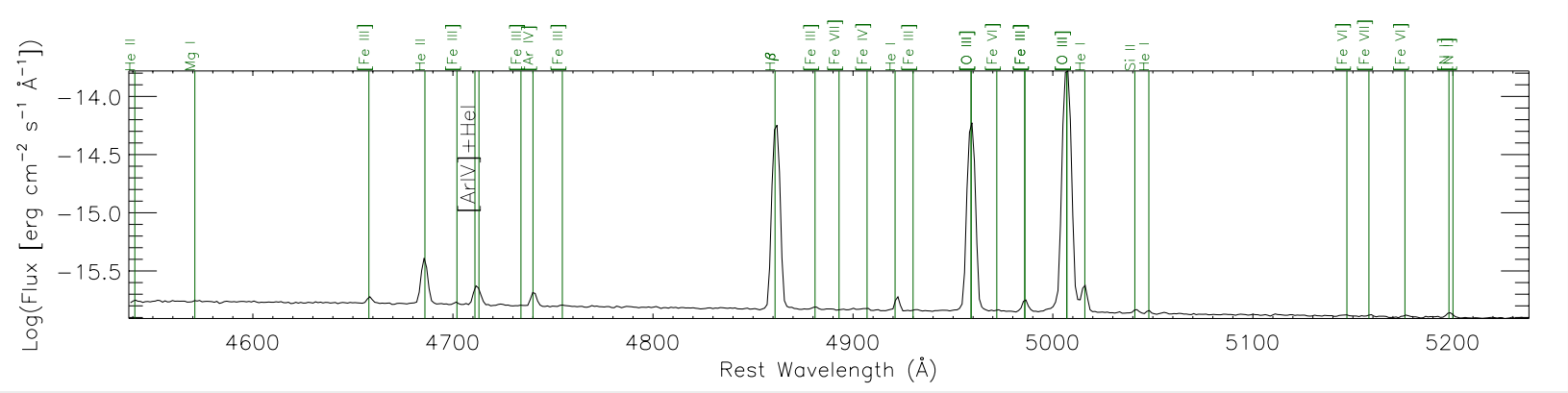}
\end{subfigure}
\begin{subfigure}
  \centering
\includegraphics[width=2\columnwidth]{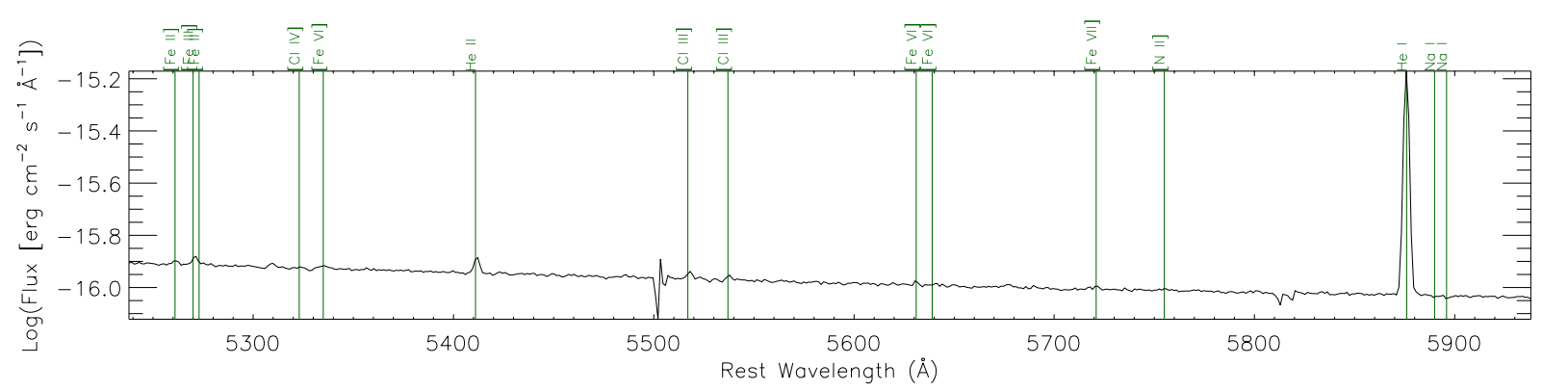}
\end{subfigure}
\begin{subfigure}
  \centering
\includegraphics[width=2\columnwidth]{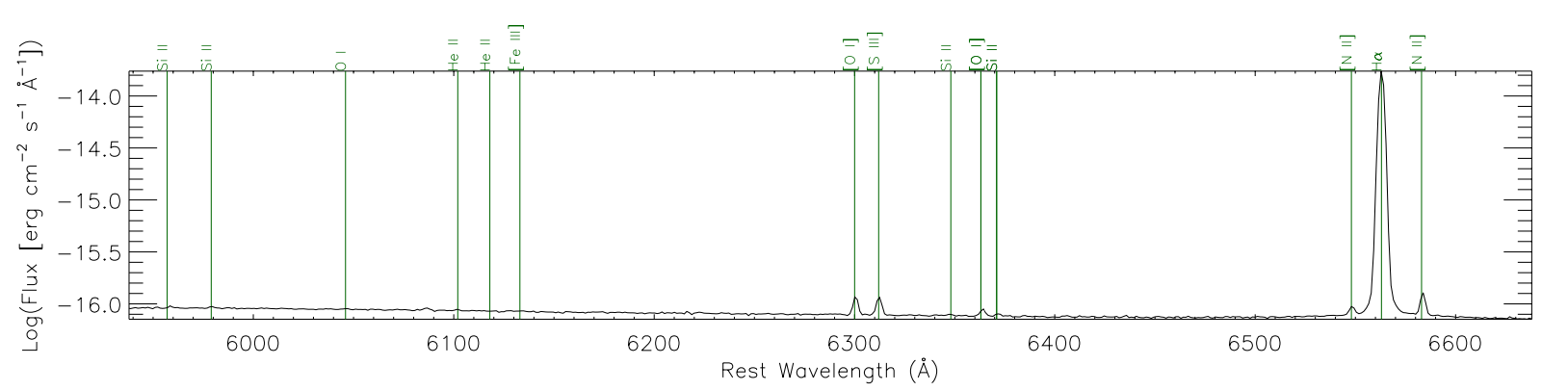}
\end{subfigure}
\begin{subfigure}
  \centering
\includegraphics[width=2\columnwidth]{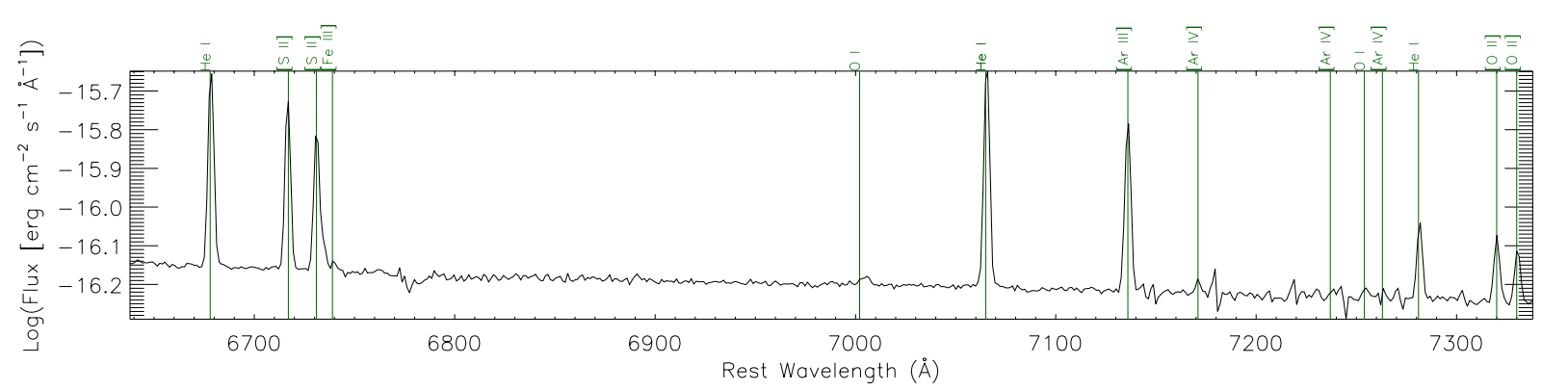}
\end{subfigure}
\begin{subfigure}
  \centering
\includegraphics[width=2\columnwidth]{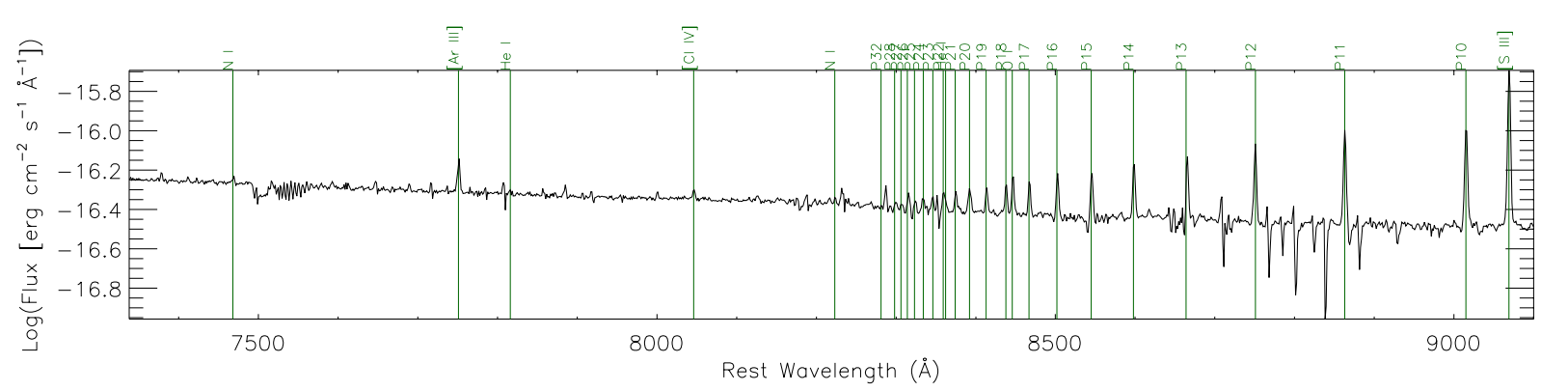}
\end{subfigure}
    \caption{MUSE spectrum of SBS 0335-052E integrated over the COS aperture. The data are plotted with a sampling of 1.23~\AA~and corrected for foreground reddening and redshift. Intrinsic spectral features are marked with solid-green lines and labeled at the top of each panel. We use the line list of \citet{Izotov2006}. Note that some of the absorption, especially in the red part of the spectrum, is caused by residuals of the sky-subtraction procedure from the pipeline.  We have, as of yet, made no attempt to correct for this. The same holds for the absorption feature at ~7500-7600 ~\AA, which is a telluric absorption band.}
\end{figure*}

In Figure~\ref{fig:muse_spectrum}, we show the MUSE spectrum which corresponds to the location and rough area of the COS aperture. The spectrum is H{\sc\,ii}-region like. Optical high-ionization emission lines of [Ar\4]$\,\lambda4740$, He\2\,$\lambda$4686 and [Ar\3]$\,\lambda$7135 are clearly detected. Also detected are the weak [N\2]\,$\lambda\lambda\,$6548, 6584 doublet around H$\alpha$ and [N\2] $\lambda$5755. The discussion of the WR features can be found in sec.~\ref{sub:wr}.

\begin{figure}\label{fig:mismatch}

\includegraphics[width=1.\columnwidth]{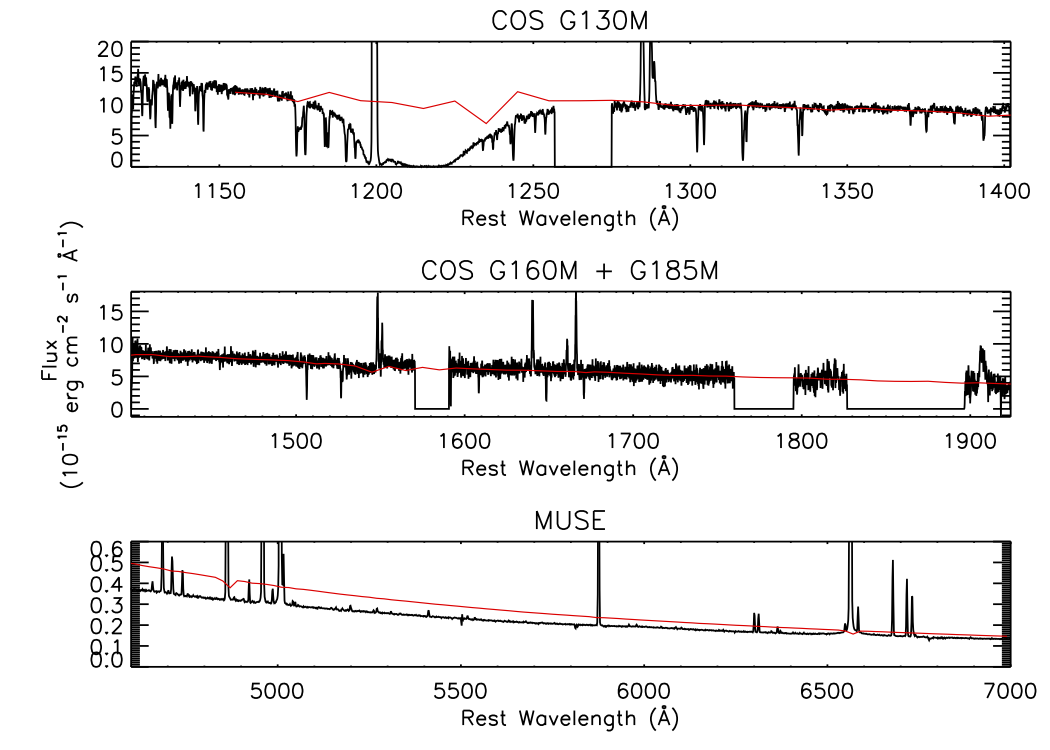}
\caption{Co-spatial COS and MUSE observations from 1120 to 7000 \AA~corrected for foreground reddening, redshift, and intrinsic reddening (black curve). We overlay a Starburst99 \citep{Leitherer2010} low spectral resolution stellar + nebular continuum model in red (see text for more details). There is a significant mismatch between the stellar + nebular continua of the COS and MUSE observations.}
\end{figure}

In order to check if the flux levels of the co-spatial COS and MUSE spectra match, we fitted a low spectral resolution stellar + nebular continuum model (SSP, $Z=0.002$, 3 Myr, non-rotating stars), which was computed with Starburst99 \citep{Leitherer2010} to these data. The observations were previously corrected for foreground reddening, redshift and intrinsic stellar reddening using the slope of the UV stellar continuum and the SMC extinction curve. The result is shown in Figure~\ref{fig:mismatch}. The model reproduces very well the continuum in the rest-frame wavelength range from 1150 to 1909 \AA~and the red portion of the optical spectrum, but shows a mismatch with the observations in the blue part of the optical spectrum. We checked that the mismatch is not due to: i) differences in coordinates between the COS and MUSE spectral extractions (they are the same); ii) the MUSE observing conditions (which were clear); iii) vignetting of COS which reduces the NUV flux (if we add UV flux, the discrepancy worsens); iv) the radius of the MUSE spectral extraction aperture; v) the reddening correction (the same correction is applied to the UV and optical); or vi) binning. Since the COS aperture is fairly large compared to the $FWHM$ of the MUSE PSF, we think that the MUSE wavelength-dependent PSF is not a big factor for the clusters near the centre of the aperture.  Unfortunately, one of the four SSCs in the COS aperture is close to the edge of the PSA making the correction for the mismatch difficult. Due to this mismatch between the optical and UV continua, we use the UV and optical separately to test our models. 

\subsection{Spectral line measurements.}\label{sub:line_measurements}

The spectra are corrected for foreground extinction using a colour excess of $E(B-V)=0.047$, which was obtained from the infrared-based Galactic dust map of \cite{Schlegel1998}, by using the reddening law (A$_\lambda$/A$_{\rm{V}}$) of Mathis (1990). They are also corrected for the redshift of the intrinsic Al\,{\sc ii}\,$\lambda$1671 line ($z=0.013475$, $v\sim4038$\,km\,s$^{-1}$). Note that we do not use the intrinsic redshift of the Si\,{\sc ii} $\lambda$1526 absorption because it is contaminated with a MW C\4 absorption. Our redshift measurement is in reasonable agreement with the average from several species in the neutral gas which was measured by \cite{James2014} using the G130M data ($z=0.013491$, $v\sim4044$\,km\,s$^{-1}$). The measurements of the intrinsic UV lines are done after the spectra have been re-binned and smoothed to match the effective spectral resolution, and corrected for foreground reddening and redshift. The measurements of the intrinsic optical lines are done after the spectra have been corrected for foreground reddening and redshift. The UV and optical line measurements are uncorrected for intrinsic attenuation due to dust.

Table~\ref{tab:centroids} gives the rest-frame wavelengths of the main lines which are analysed in this work along with the ionization potentials of the ions which are required to produce the lines. We obtained the centroid and $FWHM$ of the intrinsic nebular emission and ISM absorption lines with a custom-made IDL routine which was developed by COS science team member K. France. The routine fits a first order polynomial to the continuum and a Gaussian to each line. The radial velocities corresponding to the centroids of the lines are relative to the intrinsic Al\2\,$\lambda$1671 absorption line. The $FWHM$ and radial velocities of the intrinsic lines are reported in Table~\ref{tab:centroids}.

\begin{table}\label{tab:centroids}

\centering
\caption{Line list used in this work and their observed kinematical properties. \emph{Columns.} (1) Line ID. (2) Rest-frame wavelength in vacuum (UV lines) or air (optical lines). (3) Ion need to produce the line. (4) Ionization energy of the preceding ionization state except for the recombination He\2~$\lambda$1640 and $\lambda$4686 lines, for which we give the ionization energy of the next ionization level, which is what matters. (5) Radial velocity of intrinsic lines (relative to Al\2\,$\lambda$1671 for the UV and relative to H\al\,$\lambda$6563 in the optical. (6) Full width half maximum of intrinsic line.}
\begin{tabular}{lcccccccc}
\hline
Line ID& $\lambda_{\rm{rest}}$ & Ion & eV & $V{\rm{r}}$ & FWHM  \\
\hfill & \AA & needed & needed & km\,s$^{-1}$ & km\,s$^{-1}$ \\
(1) & (2) & (3) & (4) & (5) & (6) \\
\hline\hline
Si\2 $\lambda$1527       & 1526.71 & Si$^{+}$    & 8.15  & -2    & 92  \\
C\4 $\lambda$1548        & 1548.19 & C$^{3+}$    & 47.89 & 24    & 149  \\
C\4 $\lambda$1551        & 1550.77 & C$^{3+}$    & 47.89 & 56    & 128  \\
He\2 $\lambda$1640       & 1640.42 & He$^{2+}$   & 54.42 & -15   & 158  \\
O\3{]} $\lambda$1661     & 1660.81 & O$^{2+}$    & 35.12 & 7     & 145  \\
O\3{]} $\lambda$1666     & 1666.15 & O$^{2+}$    & 35.12 & -13   & 141  \\
Al\2 $\lambda$1671       & 1670.79 & Al$^{+}$    & 5.99 & 1      & 83  \\
{[}C\3{]} $\lambda$1907     & 1906.68 & C$^{2+}$    & 24.38 & -     & - \\
C\3{]} $\lambda$1909  & 1908.73 & C$^{2+}$    & 24.38 & -     & - \\
He\2 $\lambda$4686       & 4685.71 & He$^{2+}$   & 54.42 & 0     & 209 \\
H$\beta$  $\lambda$4861  & 4861.33 & H$^{+}$     & 13.60 & 9     & 174 \\
{[}O\3{]} $\lambda$4959  & 4958.91 & O$^{2+}$    & 35.12 & 8     & 175 \\
{[}O\3{]} $\lambda$5007  & 5006.84 & O$^{2+}$    & 35.12 & 6     & 170 \\
He\1 $\lambda$5876 & 5875.59 & He$^{+}$ & 24.6 & 13 & 144 \\
{[}O\1{]} $\lambda$6300  & 6300.30 & O${^0}$     &0&10&120 \\
{[}S\3{]} $\lambda$6310 & 6312.10 & S$^{2+}$ & 23.33 & 4 & 137 \\
{[}N\2{]} $\lambda$6548  & 6548.04 & N$^{+}$     & 14.53 & 18    & 177 \\
H$\alpha$  $\lambda$6563 & 6562.80 & H$^{+}$     & 13.60 & 10    & 137 \\
{[}N\2{]} $\lambda$6584  & 6583.46 & N$^{+}$     & 14.53 & 14    & 123 \\
{[}S\2{]} $\lambda$6717  & 6716.44 & S$^{+}$     & 10.36 & 11    & 122 \\
{[}S\2{]} $\lambda$6731  & 6731.18 & S$^{+}$     & 10.36 & 1     & 128 \\
{[}Ar\3{]} $\lambda$7135 & 7135.80 & Ar$^{2+}$   & 27.63 & 9 & 126 \\
{[}S\3{]} $\lambda$9068 & 9068.60     & S$^{2+}$ & 23.33 & 22 & 115 \\
\hline
\end{tabular}
\end{table}

\begin{figure*}\label{fig:insets}

\begin{subfigure}
 \centering
  \includegraphics[width=.32\textwidth]{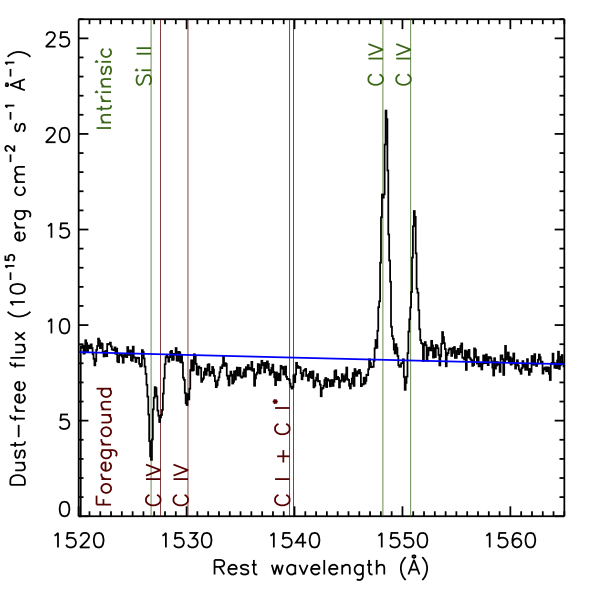}
  \end{subfigure}  
\begin{subfigure}
 \centering
  \includegraphics[width=.32\textwidth]{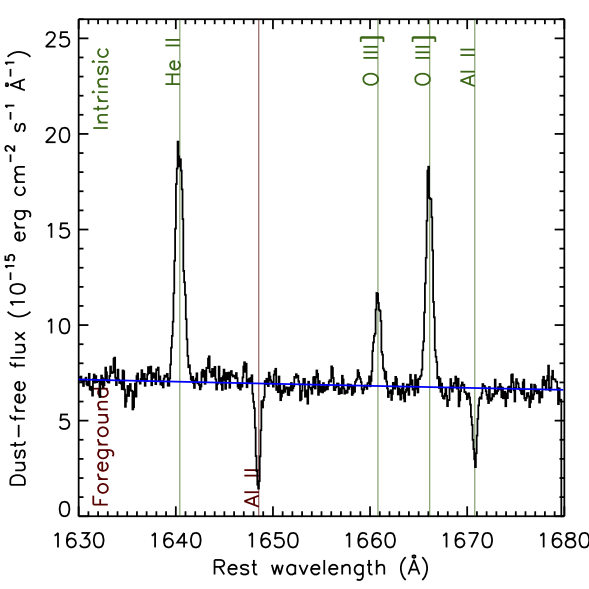}
  \end{subfigure}  
\begin{subfigure}
 \centering
  \includegraphics[width=.32\textwidth]{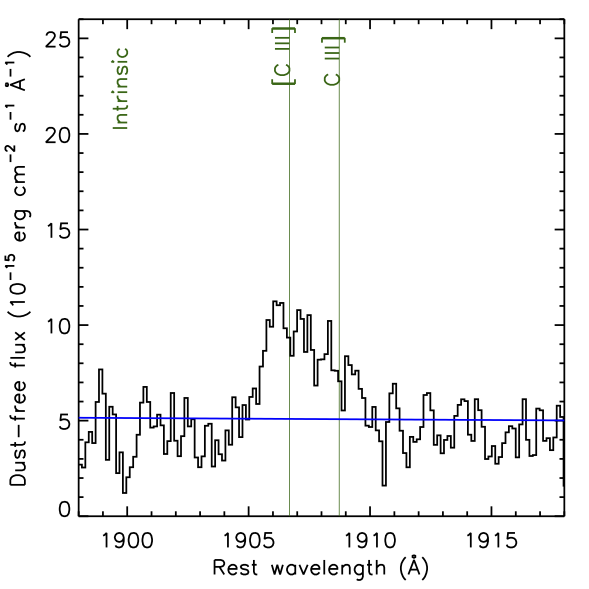}
  \end{subfigure}    
\caption{Close-up views of high-ionization UV emission lines. In the left and middle panels, we show the  weighted average of the G160M data from programs 13788 (PI Wofford) and 15193 (PI Aloisi), while in the right panel we show the G185M data from program 13788. Intrinsic lines are identified at the top, while foreground lines are identified at the bottom.  All spectra are smoothed to the effective spectral resolution ($FWHM=0.57$\,\AA) and corrected for: foreground reddening using $E(B-V)=0.047$\,mag and the Milky Way extinction law; redshift using $z_{1671}=0.01349$; and intrinsic reddening using $E(B-V)=0.017$ mag and the SMC extinction law. We overlay a power-law fit to the continuum, which is given by $F=1.72\times10^9\times\lambda^{-2.6}$ (blue curve). \emph{Left panel.} Note that the intrinsic Si\2\,$\lambda$1527 line is offset by only 4\,km\,s$^{-1}$ while the intrinsic C\4\,$\lambda\lambda$1548, 1551 nebular emission lines are redshifted by 44 and 68\,km\,s$^{-1}$, respectively.  Also note the P-Cygni like profile of C\4. \emph{Middle panel.} Note that the He\2\,$\lambda$1640 and O III]\,$\lambda\lambda$1661, 1666 nebular emission lines are at rest relative to the intrinsic Al\2\,$\lambda$1671. Also note the tentative detection of excess emission on both sides of He\2. \emph{Right panel.} Similar but for the region around [C\3], C\3]\,$\lambda\lambda$1907,\,1909. The components of the latter doublet are barely resolved.}
\end{figure*}

In order to enhance the signal-to-noise ratio (SNR) of weak components to C\4~$\lambda\lambda$1548, 1551 and He\2~$\lambda1640$, we obtained the average of the G160M data from programs 13788 (PI Wofford) and 15193 (PI Aloisi, see Table~\ref{tab:obs_log} for details), weighted by the exposure times. This was primarily done for two purposes, to better see the P-Cygni like component to C\4~$\lambda\lambda$1548, 1551, and to determine if weak broad He\2~$\lambda$1640 emission is present. Note that broad stellar He\2 emission is not necessarily expected at the metallicity of \sbs~\citep{Grafener2015}. The result is shown in the left and middle panels of Figure~\ref{fig:insets}. The C\4 $\lambda\lambda$1548, 1551 doublet is composed of a weak P-Cygni like profile and nebular-like emission which is redshifted relative to Al\2~$\lambda$1671. The redshifted C\4~lines are reminiscent of the redshifted Ly\al\ lines which are sometimes observed in Ly\al\ emitting galaxies and are explained by radiative transfer effects \cite{Verhamme2012}. He\2\,$\lambda$1640 and O\,{\sc iii}] $\lambda$1666 are in pure nebular emission ($FWHM\sim158$\,km\,s$^{-1}$) and at rest relative to Al\2 ~$\lambda$1671. Note that when comparing the profiles of He\2~$\lambda1640$ in Figure~\ref{fig:cos_spectrum} and the middle panel of Figure~\ref{fig:insets}, one can observe a slight increase in the amplitude of the He\2~which is consistent with the observational errors. No increase in the nearby continuum of the averaged spectrum is observed. 

The right panel of Figure~\ref{fig:insets} shows a close-up view of [C\3], C\3]\,$\lambda\lambda$1907,\,1909, for which only G185M data from program 13788 is available. The individual components of [C\,{\sc iii}], C\,{\sc iii}] $\lambda\lambda$1907, 1909 are barely resolved. For densities lower than or equal to $10^{3.5}\,$cm$^{-3}$, the maximum flux ratio, $F(1907)/F(1909)$, is expected to be $\sim5/3$ \citep{Ferland1981}. We find that the observed value of this ratio fluctuates around $\sim5/3$ depending on the sampling used and how we fit the continuum. 

Since the G160M data from program 15193 became publicly available when we had completed most of the analysis of the present work, and since the G160M data from our program (13788) has excellent SNR, the models tested in this work and the physical properties which we derive use the data from our program alone. The \sbs~data from program 15193 are analysed in \cite{Hernandez2020} in the context of neutral-gas chemical abundances.

Moving to the optical, the [O\3] and H$\alpha$ lines show narrow and broad components, but the flux of the broad component is negligible relative to the flux of the narrow component in both cases.  
The models which we test have a spectral sampling of 1 \AA. For comparing models with observations: i) we re-sampled the COS spectra to match the sampling of the models; and we ii) re-computed the emission line fluxes, using the coarser sampling and Trapezoidal numerical integration for profiles which were non-Gaussian. The corresponding fluxes are shown in Table~\ref{tab:numerically_int_fluxes}. For the comparison of models with observations we define the C\,{\sc iv}e flux which includes the stellar and nebular emission, and the C\,{\sc iv}a flux, which includes the stellar absorption only. Note that the spectral resolution of the MUSE data is comparable to that of the models. Thus we do not recompute the optical fluxes previous to the comparison of these with the model fluxes.

\begin{table*}\label{tab:numerically_int_fluxes}

\centering
\caption{Observed fluxes and equivalent widths of high-ionization UV emission lines. \emph{Columns}. (1) Line ID. (2) Wavelength limits of regions defining the left continuum, line, and right continuum. (3) Flux corrected for foreground reddening as described in sub-section~\ref{sub:cos}. (4) Flux error. (5) Rest-frame equivalent width. (6) Equivalent-width error. (7) Method used for obtaining the flux (T=Trapezoidal numerical integration, G=Gaussian fitting). Method T is adopted for non-Gaussian line profiles. The errors were obtained using error propagation.}
\begin{tabular}{lcccccc} 
\hline
Line ID & Wavelength ranges & $F$  & $\sigma_F$ & $EW$ & $\sigma_{EW}$ & Method \\
\hfill & \AA  & erg / s / cm$^2$ & erg / s / cm$^2$ & \AA & \AA & \hfill \\
(1)&(2)&(3)&(4)& (5) & (6) & (7) \\
\hline\hline
C\4a 1550 & 1516.4,1522.4,1540.4,1547.4,1561.4,1567.4 & -6.14E-15 & 4.58E-16 & -0.9 & 0.1 & T\\
C\4e 1550 & 1516.4,1522.4,1547.4,1552.0,1561.4,1567.4 & 1.15E-14 & 4.63E-16 & 1.7 & 0.2 & T\\
He\2 $\lambda$1640 & 1627.0,1632.0,1632.0,1650.0,1650.0,1655.0 & 1.02E-14 & 5.21E-16 & 1.7 & 0.3 & G\\
O\3{]} $\lambda$1661 & 1653.0,1659.0,1659.5,1662.1,1673.0,1677.0 & 3.90E-15 & 4.59E-16 & 0.7 & 0.1 & G\\
O\3{]} $\lambda$1666 & 1653.0,1659.0,1664.8,1667.5,1673.0,1677.0 & 9.02E-15 & 4.93E-16 & 1.5 & 0.1 & G\\
{[}C\3{]}, C\3{]} $\lambda\lambda$1907, 1909 & 1897.0, 1903.0,1903.0,1912.0,1911.0,1920.0 & 1.83E-14 & 1.35E-15 & 5.0 & 0.8 & T\\
He\2 $\lambda$4686 & 4675.0, 4680.0, 4680.0, 4690.0, 4690.0, 4695.0 & 1.62E-15 & 4.12E-17 & 4.5 & 0.0 & G \\
{[}Ar\4{]} $\lambda$4740&4720.0, 4725.0, 4737.5, 4743.5, 4745.0, 4750.0&3.89E-16&1.41E-17&1.0&0.0&G\\
H$\beta$ $\lambda$4861 & 4845.0, 4850.0, 4855.0, 4868.0, 4870.0, 4875.0 & 4.20E-14 & 8.72E-16 & 130.3 & 0.5 & G \\
{[}O\3{]} $\lambda$4959 & 4946.0, 4951.0, 4952.0, 4966.0, 4967.0, 4972.0 & 4.64E-14 & 9.60E-16 & 148.1 & 0.6 & G \\
{[}O\3{]} $\lambda$5007 & 4990.0, 4995.0, 5001.0, 5012.0, 5020.0, 5025.0 & 1.39E-13 & 2.86E-15 & 461.7 & 1.8 & G \\
He\1 $\lambda$5876 & $\lambda$5860.0, 5865.0, 5871.5, 5880.5, 5883.0, 5888.0&4.23E-15&9.23E-17&21.2&0.1&G\\
{[}O\1{]} $\lambda$6300 & 6280.0, 6290.0, 6298.0, 6303.0, 6320.0, 6330.0 & 2.90E-16 & 9.15E-18 & 1.6 & 0.0 & G \\
{[} S\3{]} $\lambda$6310&6290.0, 6295.0, 6309.5, 6315.0, 6320.0, 6325.0&2.77E-16&8.78E-18&1.6&0.0&G\\
{[}N\2{]} $\lambda$6548 & 6520.0, 6525.0, 6545.0, 6550.0, 6600.0, 6605.0 & 1.17E-16 & 6.09E-18 & 0.8 & 0.0 & G \\
H$\alpha$ $\lambda$6563 & 6520.0, 6525.0, 6552.0, 6570.0, 6600.0, 6605.0 & 1.23E-13 & 2.53E-15 & 777.2 & 3.0 & G \\
{[}N\2{]} $\lambda$6584 & 6520.0, 6525.0, 6581.0, 6587.0, 6600.0, 6605.0 & 3.40E-16 & 9.99E-18 & 2.4 & 0.0 & G \\
{[}S\2{]} $\lambda$6717 & 6708.0, 6713.0, 6713.5, 6720.5, 6721.0, 6725.0 & 8.27E-16 & 2.00E-17 & 5.3 & 0.0 & G \\
{[}S\2{]} $\lambda$6731 & 6721.0, 6725.0, 6727.5, 6734.5, 6735.0, 6740.0 & 6.65E-16 & 1.66E-17 & 4.2 & 0.0 & G \\
{[}Ar\3{]} $\lambda$7135&7120.0, 7125.0, 7132.0, 7141.0, 7150.0, 7155.0&8.22E-16&2.00E-17&6.5&0.0&G\\
{[}S\3{]} $\lambda$9068&9057.0, 9062.0, 9063.0, 9074.0, 9075.0, 9080.0&1.45E-15&3.37E-17&21.6&0.2&G\\
\hline
\end{tabular}
\end{table*}

\section{Dominant source of ionizing photons}\label{sec:ionization_source}

\cite{Izotov1997} observed SBS 0335-052E with the MMT using a 1"x181" slit. They found that the intensity of the He{\sc\,ii} $\lambda$4686 emission, integrated over the 1"x6" central region, is 3\% that of H$\beta$ and several orders of magnitude larger than the theoretical values predicted by models of photoionised H{\sc\,ii} regions. They also found that in SBS 0335-052E the peak of the He{\sc\,ii} $\lambda4686$ emission is 200 pc to the NW relative to the peak of the emission of other emission lines (H$\beta$, [O{\sc\,iii}] $\lambda$5007, 4363, and He{\sc\,i} $\lambda$5876). This led them to suggest that the He$^+$ ionization is not caused by main-sequence O stars.

\cite{Izotov2006} observed SBS 0335-052E in the spectral range $\lambda3620-9400$ \AA~at $FWHM=0.5-1\,$\AA, with the imaging spectrograph GIRAFFE on the UT2 of the VLT. Their observations cover a region of 11.4"$\times$7.3". These authors produced images of the galaxy in the continuum and in emission lines of different stages of excitation, with a spatial scale of 0.52"/pixel; and arrived to the same conclusion as \cite{Izotov1997}. Indeed, they found that while the maximum of emission in the majority of lines, including the strong lines H$\beta$ $\lambda$4861, H$\alpha$ $\lambda$6563, [O\,{\sc iii}] $\lambda\lambda$4363, 5007, [O\,{\sc iii}] $\lambda\lambda$3726, 3729, coincides with the youngest south-eastern SSCs 1 and 2, the emission of He\,{\sc ii} $\lambda$4686 line is offset to the more evolved north-west SSCs 4 and 5. This finding and the fact that the velocity dispersion of the He\,{\sc ii} $\lambda$4686 line is systematically higher, by $\sim~50-100$\%, relative to the other lines led the latter authors to suggest that the hard ionising radiation which is responsible for the He\,{\sc ii} $\lambda$4686 emission is not related to the most massive youngest stars, but rather to fast radiative shocks. In this section, we investigate if HMXB, accreting intermediate-mass black holes (IMBHs), fast-radiative shocks, WR stars, or simple stellar populations which account for massive star evolution in close binaries could be responsible for the He\,{\sc ii} $\lambda$4686 and He\,{\sc ii} $\lambda$1640 emission towards the older SSCs 4 and 5. 

\subsection{High-mass X-ray binaries.}\label{sub:hmxb}

\cite{Garnett1991} suggested that X-rays produced by massive X-ray binaries could be a source of ionization in H{\sc\,ii} regions. According to \cite{Izotov1997}, $\sim4000$ HMXBs would be required to explain the high luminosity of He{\sc\,ii} $\lambda4686$ ($5.74\times10^{38}$ erg s$^{-1}$)  in SBS 0335-052E. They obtained this number of HMXBs by directly scaling from the observed luminosity of the nebula surrounding LMC X-1, at He{\sc\,ii} $\lambda4686$, which is $1.5\times10^{36}$ erg s$^{-1}$ \citep{Pakull1986}. The equivalent number of O7 stars that these authors inferred from the H$\beta$ luminosity ($2.06\times10^{40}$ erg s$^{-1}$) is $\sim5000$. However, in a deep \textit{Chandra} observation of the galaxy, \cite{Thuan2004} detect a faint X-ray point source (29 counts, $L_X=3.5\times10^{39}$ erg s$^{-1}$) lying $\sim0.3"$ north of SSC 2. This source is consistent with only two HMXB according to table 1 of \cite{Douna2015}. Although the position of the X-ray maximum in the galaxy is no more accurate than 0.42", it is consistent with the X-ray source being physically associated with SSC 2. The timescale for HMXB formation of 3-10 Myr after the onset of the starburst is consistent with the ages of SSCs 1 and 2 (3 Myr, \citealt{Reines2008, Adamo2010}). By re-analysing the above Chandra observations, \cite{Prestwich2013} confirm that the X-ray point source which is mentioned above corresponds to an ultra-luminous X-ray source (ULX), i.e., a source having $L_{\rm{X}}>10^{39}$ erg/s. We use the coordinates of table 5 in \cite{Prestwich2013} to find the location of the ULX relative to our COS aperture. The location is indicated by the yellow circles at the southern edge of the COS footprint in Figure~1. If the ULX is responsible for the extended He{\sc\,ii} emission in \sbs, one needs to explain why the estimated location of the ULX is a few hundred parsecs away from the peak of the He{\sc\,ii} emission and completely discard other sources of ionization. 

\cite{Izotov1997} note that the elongated shapes of the faint X-ray contours in the Chandra image, to the north of SBS 0335-052E, suggest that there may also be X-ray emission  associated with SSC 3 and the pair $4+5$. There is also a very faint X-ray source to the northeast that appears to be associated with the supernova cavity seen in the optical image. Excluding the source near SSC 2, the total X-ray emission of the fainter features is $\le23$ counts, making it virtually impossible to distinguish between point-like and truly extended.

More recently, \cite{Kehrig2018} re-analized the above Chandra data. By covolving model-dependent observed X-ray luminosities with the energy-dependent cross section of He$^{+}$, they estimate the effective X-ray ionizing power. When Kehrig et al. extrapolate from the X-Ray to 4 Rydberg, they infer log Q(He\2) $\sim$ 36, while the observed optical He\2 line flux requires log Q(He\2) $\sim$ 51. This is a discrepancy of 15 orders of magnitude.  The discrepancy is not resolved, if one considers only the diffuse extended He\2. 

\subsection{Accreting IMBHs} \label{sub:imbh}

Intermediate-mass black holes ($10^2-10^6\,M_\odot$) are expected to reside in low-mass, star-forming dwarf galaxies that have not significantly grown through mergers/accretion. The detection of unresolved X-ray emission, in some cases spatially coincident with jet/core radio emission, constitutes the most compelling signature of an accreting BH in the absence of dynamical mass measurements (typically limited to the Local Group in the case of dwarf galaxies) and has provided evidence for their presence in a few tens of low-mass galaxies \citep{Mezcua2016}. There is no evidence from the combination of radio \citep{Johnson2009} and X-ray \citep{Thuan2004,Prestwich2013} data that there is an IMBH in \sbs. But where does \sbs~fall on optical diagnostic diagrams for separating star-forming galaxies from AGN?

\begin{figure*}\label{fig:bpt_diagrams}

\begin{subfigure}
  \centering
       \includegraphics[width=1.65\columnwidth]{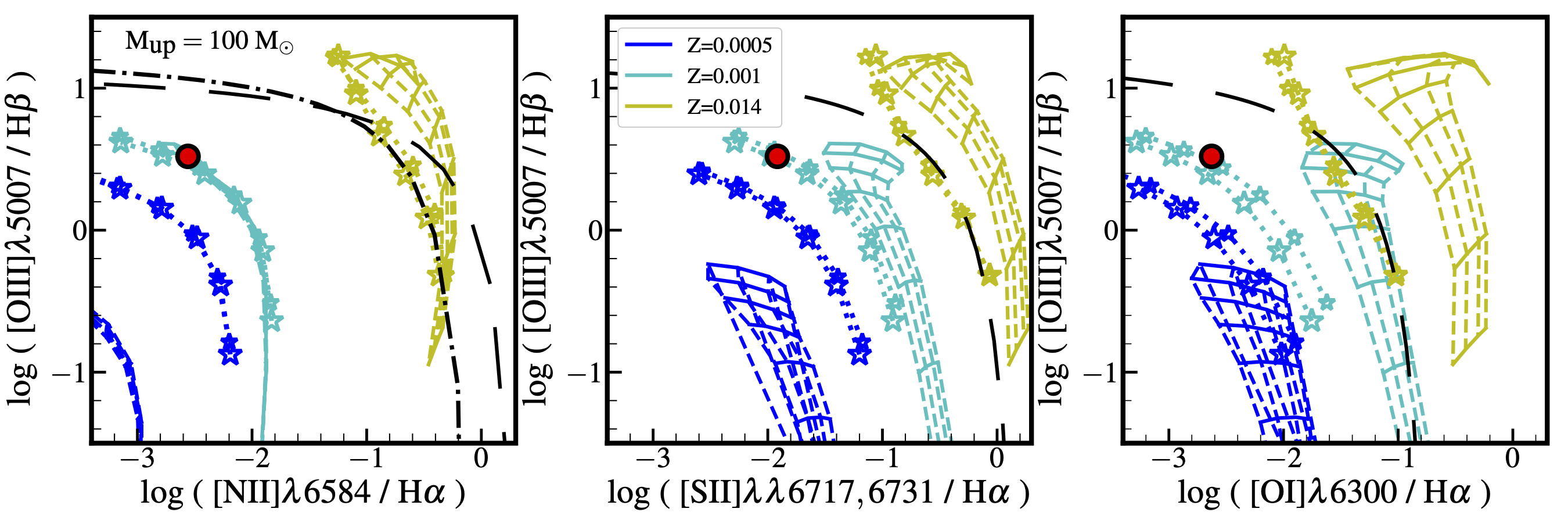}
       \includegraphics[width=1.65\columnwidth]{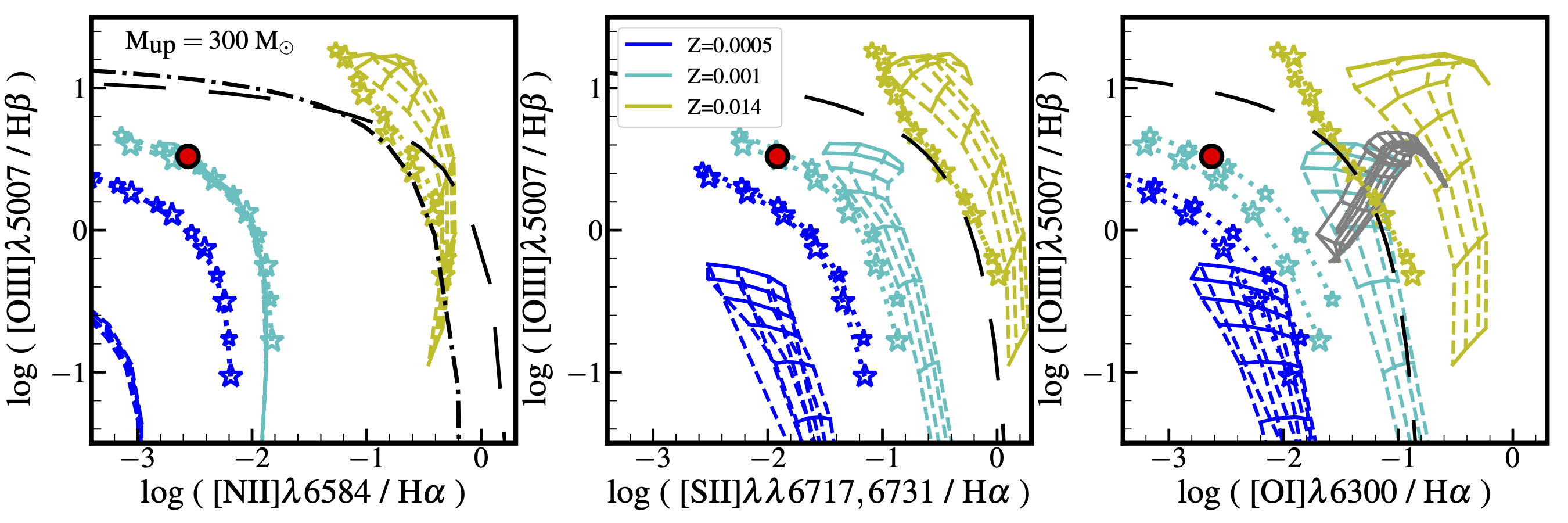}
\end{subfigure}
    \caption{Location of \sbs~(red filled circle) in the optical diagrams of \citet{Baldwin1981} and \citet{Veilleux1987} for separating narrow-line AGN from H\2~region-like galaxies. The observational errors are smaller than the symbol size. In the left-panel diagrams, the main excitation mechanism is star formation if the galaxy lies below the dotted-dashed black curve \citep{Kauffmann2003}, an AGN if the galaxy lies above the long-dashed black curve \citep{Kewley2001}, and a mixture of star formation and Seyfert nucleus, or star formation and LINER, if the galaxy lies between these curves \citep{Kewley2006}. We overlay: (i) narrow-line AGN models of \citet{Feltre2016} (dashed lines intersected by solid lines); and (ii) SSP H\2~region models of \citet{Gutkin2016} (stars connected with dotted lines). Models (i) and (ii) correspond to ionization parameters from -1.5 to -4 (H\2 regions) and -1.5 to 5 (AGN, increasing from bottom to top), three values of the interstellar metallicity (as colour-coded by the legends of the  middle panels), a dust-to-metal mass ratio of $\xi_{\rm d}=0.3$, and hydrogen gas densities of $n_{\rm H}=10^2$ (H\2 regions) and $10^3$ cm$^{-3}$ (AGN). We show H\2 region-like galaxies with: ages of $10^{6}$ and $10^{6.5}$ yr (small and big stars, respectively); IMF upper mass limits of 100 and 300 M$_{\odot}$ (top and bottom rows, respectively); and a C/O  ratio of 0.72 (C/O)$_\odot$.}
\end{figure*}

The position of \sbs~in the optical diagnostic diagrams of \cite{Baldwin1981} and \cite{Veilleux1987} is given by the red circle in Figure~\ref{fig:bpt_diagrams}. We overlay models of giant H\2 regions ionized by simple stellar populations (SSPs) computed by \citet[stars]{Gutkin2016}, and narrow-line region AGN models computed by \citet[dashed-curves]{Feltre2016}. The upper mass limit of the IMF is $M_{\rm up}=100$\,$M_\odot$ for the upper row and 300\,$M_\odot$ for the bottom row. There are no significant differences between the H\2-region models of the top and bottom panels. Most importantly, at extreme-low metallicity (dark-blue and cyan models), the AGN models (dashed lines) are located on the star-forming zone set by \cite{Kewley2001}. Furthermore, the AGN and star-forming models overlap in the first diagnostic diagram (first column) at Z=0.001, and the last diagnostic diagram (last column) at Z=0.0005. Thus, at the metallicities of XMPs, these optical diagrams are not useful for separating AGN from star-forming regions. Note that the SSP + H\2 region models of $Z=0.001$ approach the \sbs~observation in the three optical diagnostic diagrams.

\begin{figure*}\label{fig:SBS_UV}

\includegraphics[width=1.35\columnwidth]{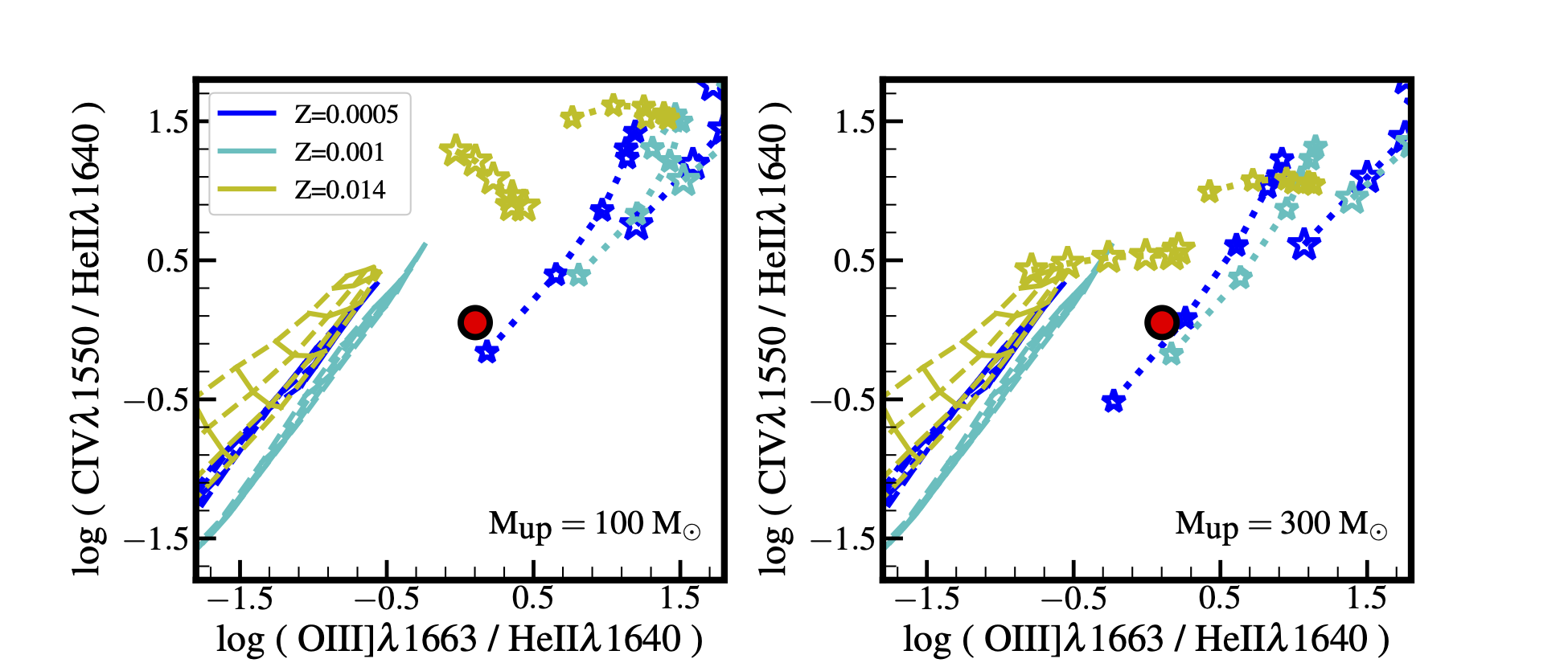}
    \caption{Location of \sbs~(red filled circle) in the UV diagnostic diagram, C\,{\sc iv} ($\lambda$1549\,+\,$\lambda$1551)\,/\,He\,{\sc ii} $\lambda$$1640$ versus O\,{\sc iii}] ($\lambda$1661\,+\,$\lambda$1666)\,/\,He\,{\sc ii} $\lambda$1640. The observational error bars are smaller than the symbol size. We use the same models and colour scheme as in Figure \ref{fig:bpt_diagrams}. The upper mass limit of the stellar IMF is 100 M$_{\odot}$ in the left panel and 300 M$_{\odot}$ in the right panel.}
    
\end{figure*}

\cite{Feltre2016}, computed predictions of high-ionization UV emission lines from narrow line region models of AGN and proposed diagnostic diagrams for separating AGN from star-forming galaxies which are based on UV emission-line ratios. These  diagrams separate better star-forming galaxies from AGN at extreme low metallicities, as can be seen in the diagrams which are shown in Figure~\ref{fig:SBS_UV}. In the latter figure, the left and right panels show SSPs with $M_{\rm up}=100$\,$M_\odot$ and 300 $M_\odot$, respectively. Although the association of blueshifted O\5\,$\lambda$1371 with VMSs is not definitive, models with VMSs are more likely to reproduce the fluxes of UV high-ionization lines than models with an upper mass limit of 100 $M_\odot$ \citep{Gutkin2016}, which is why we test them in this work. Only the Z=0.0005 and $M_{\rm up}=300$\,$M_\odot$ SSP + H\2~region model approaches the UV line ratios of \sbs. However, the closest model has nebular + broad He\2~$\lambda1640$ emission. As already mentioned, broad He\2 emission is not necessarily expected at the low metallicity of \sbs, and in this galaxy it is not observed (see Figure~\ref{fig:insets}).  

\begin{figure}\label{fig:spitzer_cc}

\centering
\includegraphics[width=0.69\columnwidth]{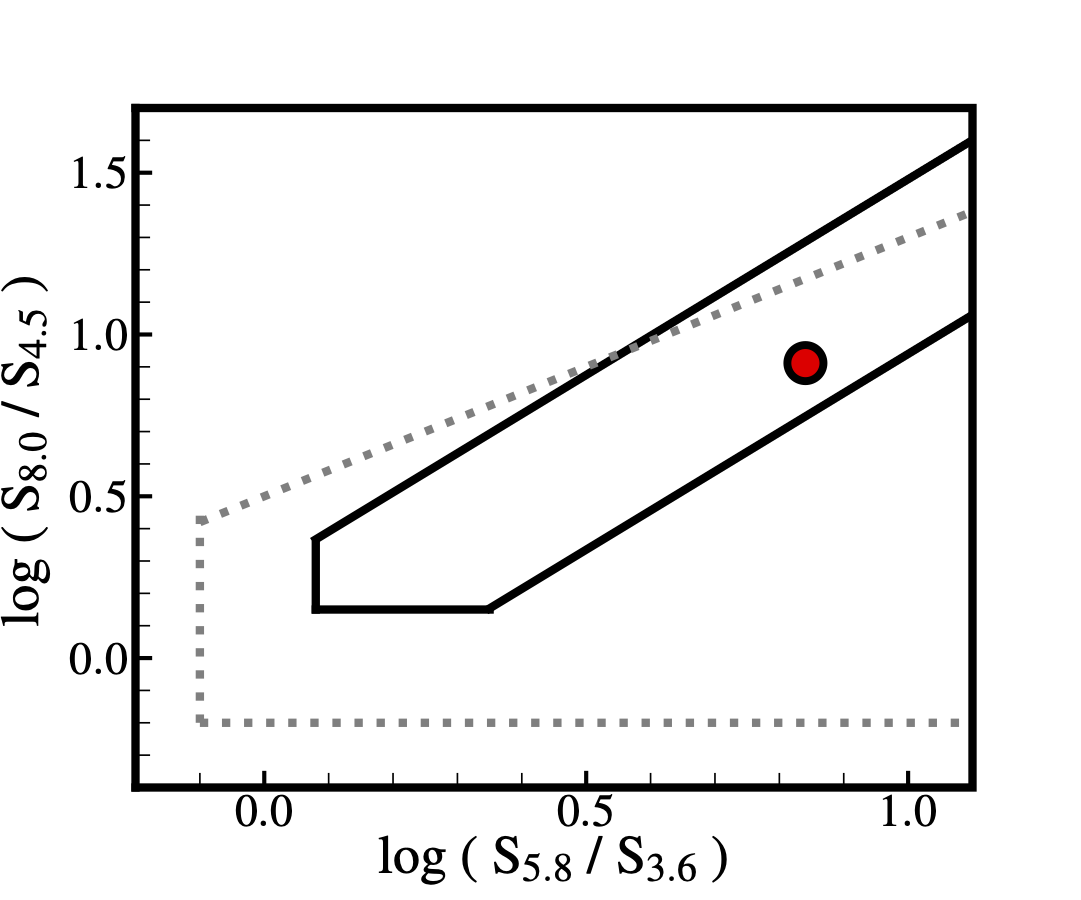}
\caption{Wedges of Lacy et al., 2004 and Donley et al., 2012 (dashed gray and continuous black lines, respectively) that select AGN in the \textit{Spitzer}/IRAC colour-colour diagram. The non-dashed line represents a more strict cut. SBS 0335-052E (red symbol) would be selected as an AGN based on this diagram. We use the data from the IRAC 3.8`` aperture which is provided by NASA/IPAC Extragalactic Database (NED).}
\end{figure}

We now discuss some of the IR properties of ~\sbs~in the context of this subsection. In Figure~\ref{fig:spitzer_cc} we show that~\sbs~falls within the wedges of \cite{Lacy2004} and \cite{Donley2012} that select AGN in the \textit{Spitzer}/IRAC colour-colour diagram. The position of \sbs~in Figure~\ref{fig:spitzer_cc} is attributed to an IR excess due to warm dust. Figure 1 of \cite{Houck2004} shows the Spitzer IRS spectrum of \sbs. The latter authors say that if the mid-IR spectrum of \sbs~is representative of star-forming galaxies at higher redshifts, it may be difficult to distinguish between galaxies whose emission is dominated by star-formation and AGN, as the latter case is associated with a relatively featureless flat spectra in the mid-IR. 

\subsection{Fast radiative shocks from SNRs}\label{sub:shocks}

\cite{Garnett1991} suggested that radiative shocks produce strong He{\sc\,ii} emission under certain conditions. The strength of such emission is sensitive mostly to the velocity of the shock, reaching a maximum for $V_{\rm{shock}}\sim120$ km s$^{-1}$, dropping rapidly at higher velocities. 

The \textit{FUSE} spectrum of SBS 0335-052E which is presented in \cite{Grimes2009} shows no O{\sc\,vi} in emission at 1032-1038 \AA, which at higher metallicities is generally interpreted as an indication of the presence of shocked gas from SNRs \citep{Rasmussen1992, Allen2008}. For reference, the size of the \textit{FUSE} aperture is 25"x25". It is not clear if the absence of O{\sc\,vi} in emission is due to a combination of the low metallicity of the galaxy and the strong nearby Ly$\beta$ absorption.

\cite{Kehrig2018}, analysed the MUSE spectra of~\sbs~but did not correct for the fact that the MUSE DRS sky-subtraction algorithm initially oversubtracted H$\alpha$ and [O\3] emission (see our Section~\ref{sub:muse_obs}). Thus, we expect their flux measurements of the above lines to suffer from this oversubtraction. They find no high ratios of [S\2]/H\al~and/or [O\1]/H\al~in \sbs. At higher metallicities, high values of the latter ratios indicate the presence of shock-excited gas. 

Figure 6 of~\cite{Alarie2019}, shows optical emission line ratios for shock + precursor models of extremely low metallicity in the right panel. Note that what is actually plotted in that figure is log([OIII]5007/H$\beta$) and  log([NII]6584/H$\alpha$) (private communication). At SMC metallicity, the models of~\cite{Alarie2019} are in excellent agreement with the similar models of \cite{Allen2008}. However, SMC metallicity is the lowest metallicity which is available for the models of \cite{Allen2008} and this is higher than the known metallicity of~\sbs. Thus, we compare against the models of~\cite{Alarie2019}. We find that the optical ratios of~\sbs, log([OIII]5007/H$\beta$)=0.52 and log([NII]6584/H$\alpha$)=-2.56, are incompatible with the values of the extremely low metallicity shock + precursor models of \cite{Alarie2019}. 

\subsection{Wolf-Rayet stars.}\label{sub:wr} 

Wolf-Rayet (WR) stars are the evolved descendants of stars with initial masses of $>25-30\,M_\odot$. They are divided into two classes, those with strong lines of He and N (WN class) and those with strong He, C, and O (WC and WO class). Depending on the relative intensities of their N or C lines, these classes are further divided into subtypes. WN spectral subtypes follow a scheme involving line ratios of N\3-\5 and He\1-\2, ranging from WN2 to WN5 for early WN (WNE) stars, and WN7 to WN9 for late WN (WNL) stars. WC spectral subtypes depend on the line ratios of C\3 and C\4 lines along with the appearance of O\3-\5, spanning WC4 to WC9 subtypes, for which WC$4-6$ stars are early (WCE) and WC$7-9$ are late (WCL). Rare, O-rich WO stars form an extension of the WCE sequence, exhibiting strong O\6 $\lambda\lambda$ $3811-34$ emission. A review of the spectral and physical properties of these stars is provided in \cite{Crowther2007}. 

\cite{Schaerer1998} modelled the nebular and WR He{\sc\,ii} $\lambda4686$ emission in young starbursts with metallicities between 0.2 and 1 times $Z_\odot$, and predicted a strong nebular He{\sc\,ii} $\lambda4686$ emission due to a significant fraction of WC stars in the early WR phases of the burst. Based on the fact that observations of extreme-low metallicity young starbursts lack broad WR-like emission features, \cite{Izotov1997} excluded WR stars as the explanation for the intense He{\sc\,ii} $\lambda4686$ emission in SBS 0335-052E. However, more recent works have reported tentative detections of WR stars in SSC 3, which is within the COS aperture. 

\cite{Papaderos2006} observed the galaxy with a 1.2" wide slit at the 3.6 ESO telescope. They report the tentative detection of WC4 (early carbon-type WR) stars in SSC 3, which is within our COS aperture. The evidence is a broad emission feature which is coincident with the [Fe\3] $\lambda4658$ nebular emission line but has a width much larger than those of other nebular emission lines in the spectrum of SSC \#3. \citep{Papaderos2006}  interpret this broad feature as due to C\4 $\lambda$4658, which is a signature of WC4 stars. From this feature, corrected for [Fe\3] $\lambda4658$ emission, they estimate a total number of $\sim$22 WC4 stars. Given the derived number of O7V stars ($\sim800$), the ratio of WC4 to O7V stars they find is consistent with the stellar population synthesis models at solar metallicity. \cite{Papaderos2006} point out that  \cite{Crowther2007} showed that the line luminosities of extremely low-metallicity WC4 stars are 3-6 times lower than those predicted by the above models, which would increase the number of WC4 stars to 70-130. However, the latter authors say that in order to confirm the detection of WC4 stars and to better determine their number, better observations of the C\4 $\lambda$5808 emission line are required. Finally,  \cite{Papaderos2006} found: weak N\3 $\lambda4640$ and He\2 $\lambda4686$ emission in SSC 3, thus the determination of the number of WNL stars was not possible; and no WR features in regions 1, 2, 4, and 5. 

\begin{figure*}\label{fig:wr}

\includegraphics[width=1.5\columnwidth]{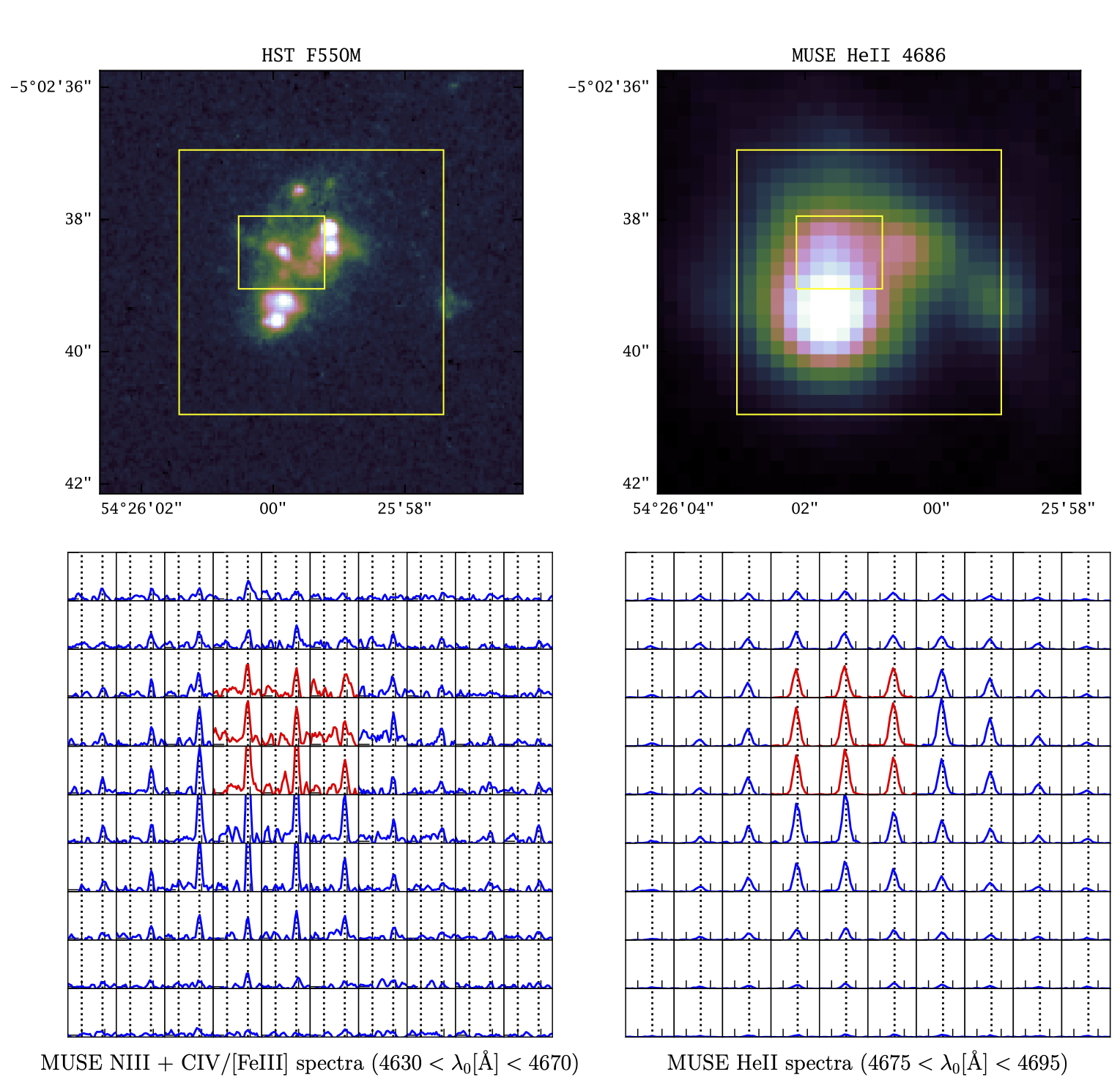} 
\caption{Tentative evidence for broad WR emission lines N{\sc\,iii} $\lambda4640$ and C{\sc\,iv}/[Fe{\sc\,iii}] $\lambda$4658 in MUSE observations. \emph{Top
left}: {\it{HST}} F550M image of SBS 0335-052E.  The spectral window of this medium band filter does not coincide with any emission lines from the galaxy, therefore it represents an ideal representation of the stellar light.  The big yellow box represents the 4 arcsec $\times$ 4 arcsec region for which the spectra from the MUSE cube (shown in the bottom panels) have been extracted, while the small yellow box indicates a 1.2 arcsec $\times$ 1.2 arcsec region around SSC 3. \emph{Top right:} Continuum subtracted He\2 $\lambda4686$ image from MUSE data shown for the same field of view as in the top right panel. The image is displayed using an arcsinh-stretch, with the brightest pixels corresponding to flux densities of $
10^{-16}$erg\,s$^{-1}$cm$^{-2}$arcsec$^{-2}$.  \emph{Bottom left}:
MUSE continuum subtracted spectra, each extracted over 2$\times$2
spaxels (i.e. quadratic apertures of 0.4 arcsec\,$\times$\,0.4
arcsec), showing the region around the broad WR signatures
NIII $\lambda4640$ and C{\sc\,iv} $\lambda$4658 (indicated by vertical dotted
lines).  The C{\sc\,iv} $\lambda$4658 emission feature is blended with nebular
[Fe{\sc\,iii}] $\lambda4658$ emission.  Shown in red are the spectra tracing
emission from cluster 3, i.e. within the region outlined by the
smaller yellow box in the top panels.  The scale on the y-axis is
linear from 0 to 8$\times10^{-18}$erg\,s$^{-1}$\AA{}$^{-1}$ and the
spectra have been smoothed by a 3.6~\AA{} (3 spectral pixel) wide
boxcar filter in order to enhance the visibility of the faint
features.  \emph{Bottom right}: Same as the bottom left panel, but for
He{\sc\,ii} $\lambda4686$ emission with the y-axis extending from 0 to
$10^{-16}$ erg\,s$^{-1}$\,\AA{}$^{-1}$ and with unsmoothed spectra.}
\end{figure*}

We used the MUSE spectra to further investigate the presence of WR stars. For this purpose, we extracted a smaller datacube around SSC 3. The results are shown in Figure~\ref{fig:wr}. The data show that WC4 stars could be present in the vicinity of SSC 3. However, we find no evidence of C\4\,$\lambda$5808 emission in the vicinity of this SSC. As previously mentioned, \cite{Kehrig2018} independently analysed the MUSE data (uncorrected for the initial MUSE DRS sky-oversubtraction). They found a WR knot in \sbs. However, their analysis of the data discards single WR stars as the dominant source of He\2 ionization. 

\subsection{SSPs with massive stars in close-binary systems}\label{sub:bpass}

There is observational evidence that massive stars are in binary systems with close to 70\% interacting over the course of their evolution \citep{Massey2009, Chini2012, Sana2012, Sana2013}. \texttt{BPASS v2.1} SSP models which account for massive star evolution in close binaries are presented in \citet[stellar component]{Eldridge2017} and \citet[nebular component]{Xiao2018}. Note that at the metallicity of \sbs~($Z\sim 6\times10^{-4}$), there are no observational constraints for individual close-binary systems.

\cite{Kehrig2018} used models which account for single rotating stars and binaries to interpret the integrated He\2 $\lambda$4686 luminosity of \sbs~and found that it can only be reproduced with single rotating stars if the stars are metal-free and with binaries if the IMF is top-heavy and the stellar metallicity is  $Z\sim10^{-5}$. Figure 12 of \cite{Stanway2019} confirms that \texttt{BPASS v2.1} binary models with a metallicity of $Z=10^{-5}$ can reproduce the integrated He\2$\,\lambda$4686 luminosity of~\sbs, which is reported in \cite{Kehrig2018}. However, $Z=10^{-5}$ is unrealistically-low for~\sbs.

\begin{figure*}\label{fig:bpass_plots}

\includegraphics[width=1.49\columnwidth]{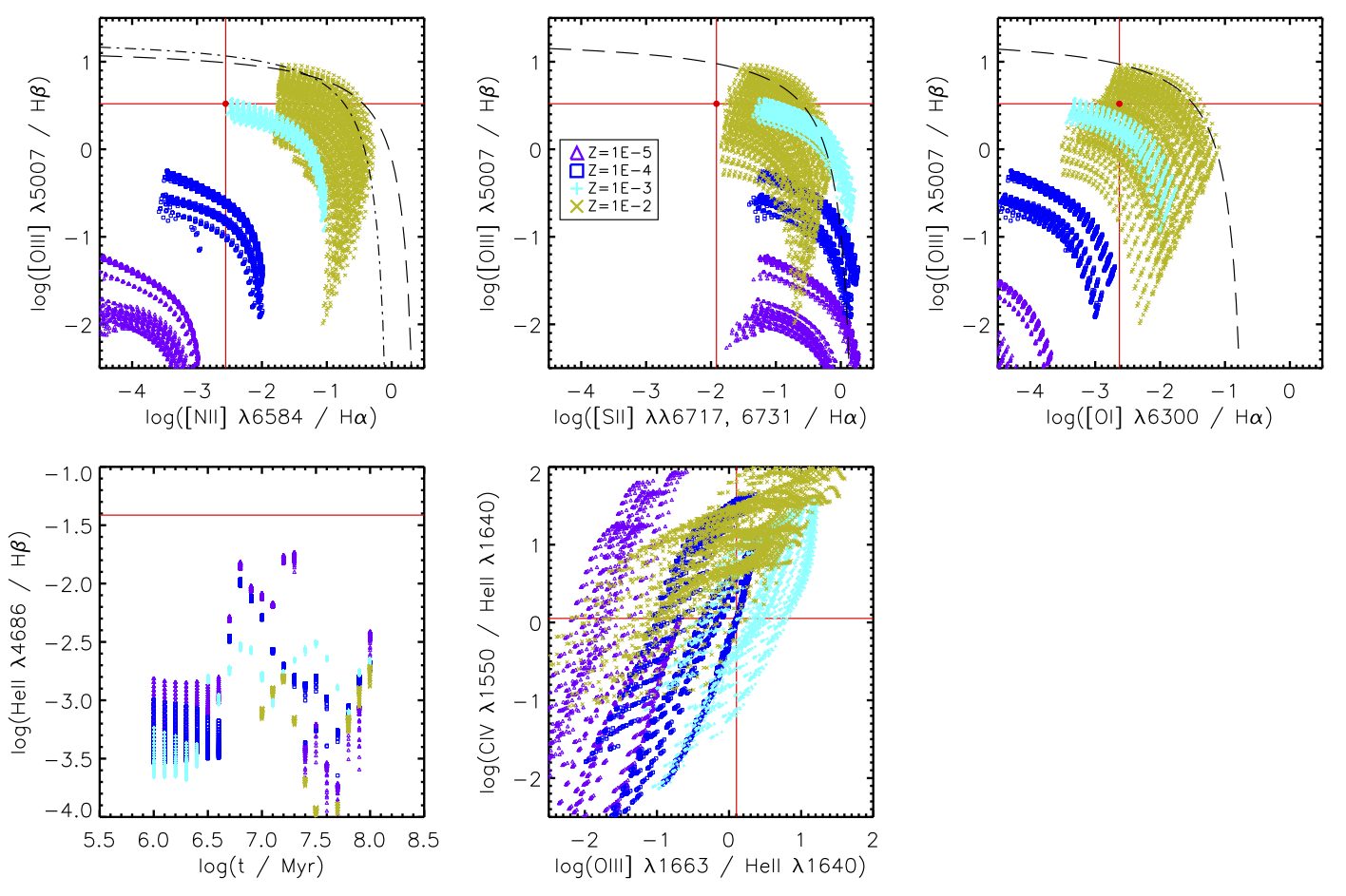}
\caption{\emph{Top row.} Same diagrams as in Figure~\ref{fig:bpt_diagrams} but we show the location of \sbs~(red filled circle at intersection of vertical and horizontal red lines) relative to the SSP \texttt{BPASS v2.1} models which are described in subsection~\ref{sub:bpass}. The radius of the red circle is larger than or equal to the observational error bar. \emph{Bottom-left.} Location of \sbs~(horizontal red line) relative to \texttt{BPASS v2.1} models in the He\2$\,\lambda$4686/H$\beta$ versus age diagram. The y-axis error bar is of the size of the thickness of the red vertical line. \emph{Bottom-middle.} Location of \sbs~(red filled circle at intersection of vertical and horizontal red lines) relative to \texttt{BPASS v2.1} models in the UV diagnostic diagram. We show models of $M_{up}=300\,M_\odot$ and metallicities given in the legend of the top-middle panel. We include models with log(U) values ranging from -3.5 to -1.5 in steps of -0.5 and log(n[H]) values ranging from 0 to 3 in steps of 1.}
\end{figure*}

Here we investigate if the \texttt{BPASS v2.1} binary models with nebular emission can reproduce the optical and UV emission line ratios which we measured for~\sbs.  Figure~\ref{fig:bpass_plots} shows optical and UV diagnostic diagrams similar to those of Figures~\ref{fig:bpt_diagrams} (bottom row) and~\ref{fig:SBS_UV} (right panel). In addition, in the bottom left panel of Figure~\ref{fig:bpass_plots} we show a plot of He\2\,$\lambda4686/$H$\beta$ versus age.  In Figure~\ref{fig:bpass_plots}, the position of \sbs~(red filled symbol) is shown relative to \texttt{BPASS v2.1} models. 
 
We find that that the binary models which are closest to the observations are of different metallicities in each diagram. In particular, the top-right panel shows that only models with a metallicity higher than that of~\sbs~can reproduce the observed emission line ratios. Furthermore, none of the binary models are able to reproduce the observed He\2 $\lambda$4686/H$\beta$ ratio, as can be seen in the bottom-left panel of Figure~\ref{fig:bpass_plots}. On the other hand, some of the binary models with $Z=0.001$ approach the UV emission line ratios of~\sbs, as shown in the bottom-right panel of the same Figure. Thus, we conclude that the SSP + H II region \texttt{BPASS v2.1} binary models fail to reproduce the optical emission line ratios of~\sbs.

\section{Comparison of models with observations}\label{sec:models_tool_tests}

In this section, we present the models and spectral interpretation tool which we use in this work. We also  present results from five tests where we compare the models to different sets of observables.

\subsection{Models and Tool}\label{sub:models_tool}

We compare models to the co-spatial COS and MUSE observations separately. For this purpose, we use the spectral interpretation tool, \beagle\ \citep{Chevallard2016}, which incorporates in a flexible and consistent way the production of radiation from stars and its transfer through the interstellar medium. The version of \beagle\ which we use relies on the models of \cite{Gutkin2016}, who follow the prescription of \cite{Charlot2001} to describe the emission from stars and the interstellar gas. In particular, the models are computed combining the latest version of the \citet{Bruzual2003} stellar population synthesis model with the standard photoionization code CLOUDY \citep{ferland2013}. The latest implementation of \citet{Bruzual2003} use the most up-to-date single non-rotating massive-star evolution models from the Padova group \citep{Tang2014}.

Using single non-rotating models, as we do in the present work, sets a reference for comparison with more comprehensive models. Also note that before interacting and after merging, massive stars evolve as single stars.  The ionized-gas O/H and C/O which are obtained via standard methods disagree with the metal-free or unrealistically-low metallicity scenarios.

In the models of \cite{Gutkin2016}, the main adjustable parameters of the photoionized gas are the interstellar metallicity, \Zism; the typical ionization parameter of newly ionized \Hii\ regions, \Us, which characterizes the ratio of ionizing-photon to gas densities at the edge of the Str\"omgren sphere; and the dust-to-metal mass ratio, \xid, which characterizes the depletion of metals on to dust grains. We let the C/O number abundance ratio vary, but only consider models with hydrogen density $\nH=100\,\mathrm{cm}^{-3}$. In their figures 5 (all panels) and 10 (panels b and d), \cite{Gutkin2016} show that at low metallicity, changing $\nH$ from 10 to 1000$\,\mathrm{cm}^{-3}$ has a negligible effect on the ratios of the optical and UV lines which are used in our paper. For the attenuation by dust, we adopt the 2-component model of \citet{Charlot2000} to account for the effects linked to dust/star geometry (including ISM clumpiness). Table~\ref{tab:priors} gives the ranges of the main adjustable parameters. In the table, \tauV\ is the $V$-band dust attenuation optical depth and $\mu$ the fraction of Fraction of $\hat\tau_V$ arising from the diffuse ISM rather than from giant molecular clouds. We adopt the same metallicity for stars and star-forming gas ($Z_{\rm stars}=Z_{\rm ISM}$) and assume that all stars in a galaxy have the same metallicity. By fitting the \hst~photometry of the four SSCs which are located within the COS aperture, \cite{Reines2008} and \cite{ Adamo2010} find an age range of 6 to 12 Myr. Given this age range, in this work, we test models of constant star formation. The adopted ranges for the galaxy mass, log(mass/$M_\odot$), and time since the beginning of star formation, log(t/yr), are provided in Table~\ref{tab:priors}. We adopt a \citet{Chabrier2003} IMF with an upper mass limit of $M_{\rm up}=300$ \msun, i.e., we account for the presence of VMSs, given that we found a blueshifted O\5 1371 absorption. 

With this parameterization, we use \beagle~to fit the available observational constraints. Note that we do not fit any observables which depend on the velocity dispersion of the emission lines. As output, we obtain the posterior probability distributions of the above adjustable model parameters. By fitting the UV and optical data independently, we obtain independent measures of the time since the beginning of star formation and the gas metallicity. 

\begin{table}\label{tab:priors}

\centering
\caption{Grid sampling of main adjustable parameters of photoionization models which are described in Section~\ref{sub:models_tool}}
\begin{tabular}{ll} 
\hline
Parameter & Sampled values \\
\hline\hline
    log(\Zism/$Z_\odot$)$^{\rm{a}}$ & -2.2 to 0.24\\
    log(U$_{\rm S}$) & -4.0, -3.5, -3.0, -2.5, -2.0, -1.5, -1.0 \\
    $\xi_{\rm d}$ & 0.1, 0.3, 0.5 \\  
    $\hat\tau_V$ & 0 to 5 \\
    $\mu$ & 0 to 1 \\        
    $\nH$ ($\mathrm{cm}^{-3}$) & 100 \\
    $(\mathrm{C/O})/(\mathrm{C/O})_\odot$$^{\rm{b}}$ & 0.14, 0.52, 1.0 \\
    log(t/yr)$^{\rm{c}}$ & 6.0 to 10.0 \\
    log(mass/$M_\odot$)$^{\rm{d}}$ & 3.0 to 11.0\\
\hline
\end{tabular}
\begin{tablenotes}
      \item \textbf{Notes} \\
      \item $^{\rm{a}}$ $Z_\odot=0.01524$. See \cite{Gutkin2016} for metallicity steps.      
      \item $^{\rm{b}}$ (C/O)$_\odot=0.44$.
      \item $^{\rm{c}}$ Time since the beginning of star formation (max\_stellar\_age in \beagle).
      \item $^{\rm{d}}$ M$_\odot=1.989\times10^{33}$\,g.
    \end{tablenotes}    
\end{table}

\subsection{Test 1: fit to P-Cygni\,+\,nebular C\4~1550 profile}\label{sub:test1}

When the integrated spectrum of a star-forming galaxy shows a P-Cygni like C\4~$\lambda\lambda\,$1548, 1551  profile, the profile is generally interpreted as originating in the winds of O-type MS stars \citep[section 3.4]{Wofford2011}. Figure 24 of \cite{Leitherer2010} shows the C\4~profile of a simple stellar population. The C\4~emission component is strongest a few Myr after the burst of star-formation, when O-type MS stars are still present. Figure 29 of \cite{Leitherer2010} shows the C\4~profile in the case of constant star formation, 50 Myr after the beginning of star formation, i.e., when there is an equilibrium between birth and death of stars. In this case, C\4~still shows a P-Cygni like profile. Figure 29 of \cite{Leitherer2010} also shows that the C\4~absorption component is sensitive to the metallicity of the stellar population.  Figures 16-19 of \cite{Vidal-Garcia2017} show similar behaviours of C\4~for the models which are used in the present work. 

\begin{figure}\label{fig:test1_pix2pix}

\centering
\includegraphics[width=0.89\columnwidth]{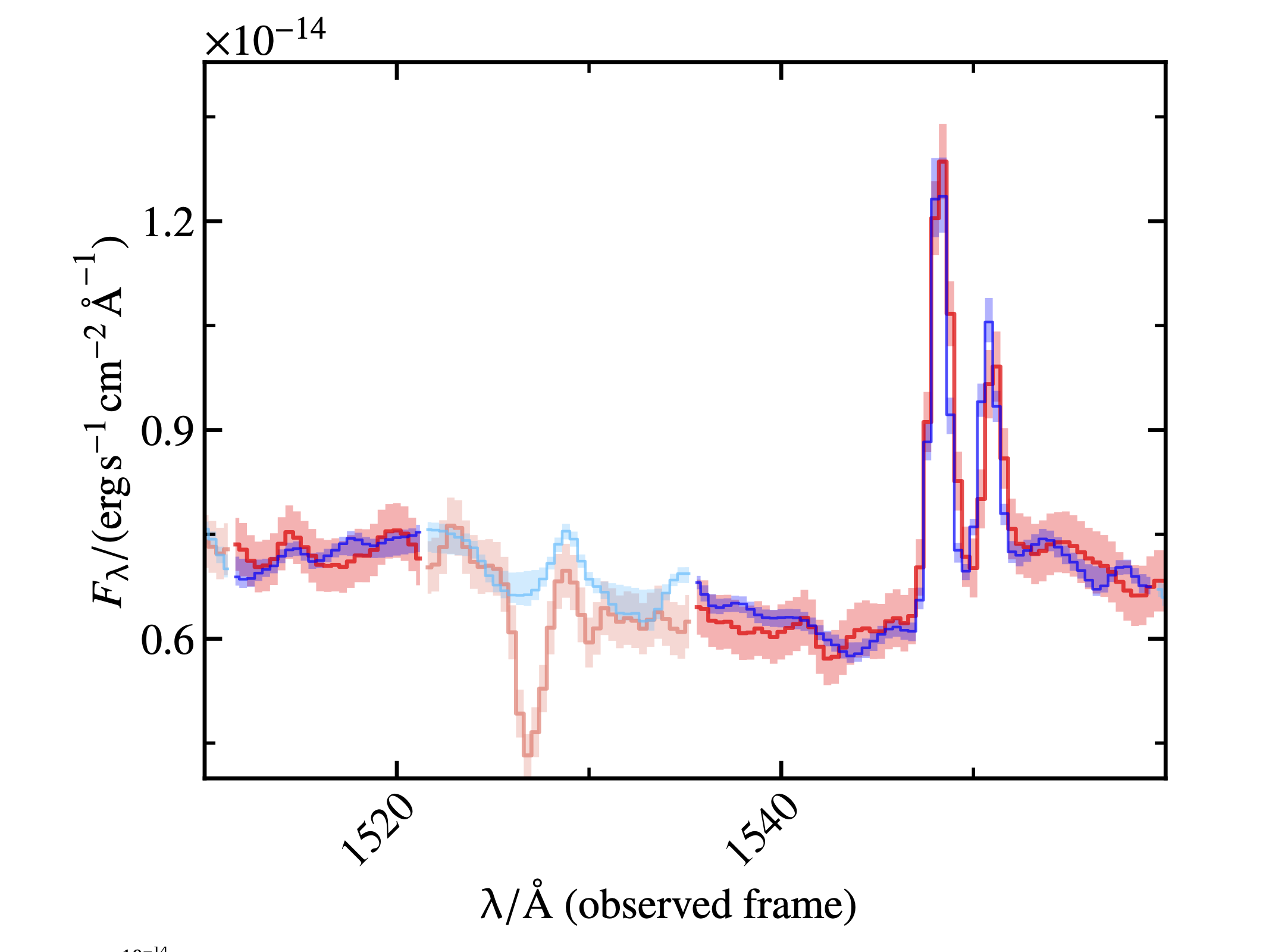}
\caption{Fit to observed C\4\,$\lambda\lambda$1548, 1551 P-Cygni like + nebular profile. The red curve and shadow correspond to the observations and observational uncertainties, respectively. The dark-blue curve and shadow represents the posterior median of models and 68\% posterior credible interval, respectively. The cyan portion of the spectrum was excluded from the fit as it contains interstellar absorptions which are not accounted for in the models.}
\end{figure}

As test 1, we attempt to fit the P-Cygni like plus nebular C\4~profile of \sbs~and nothing else. Figure~\ref{fig:test1_pix2pix} shows that the models are able to reproduce the combined stellar + nebular profile. 

\begin{figure}\label{fig:test1_triangle}

 \centering
  \includegraphics[width=0.69\columnwidth]
  {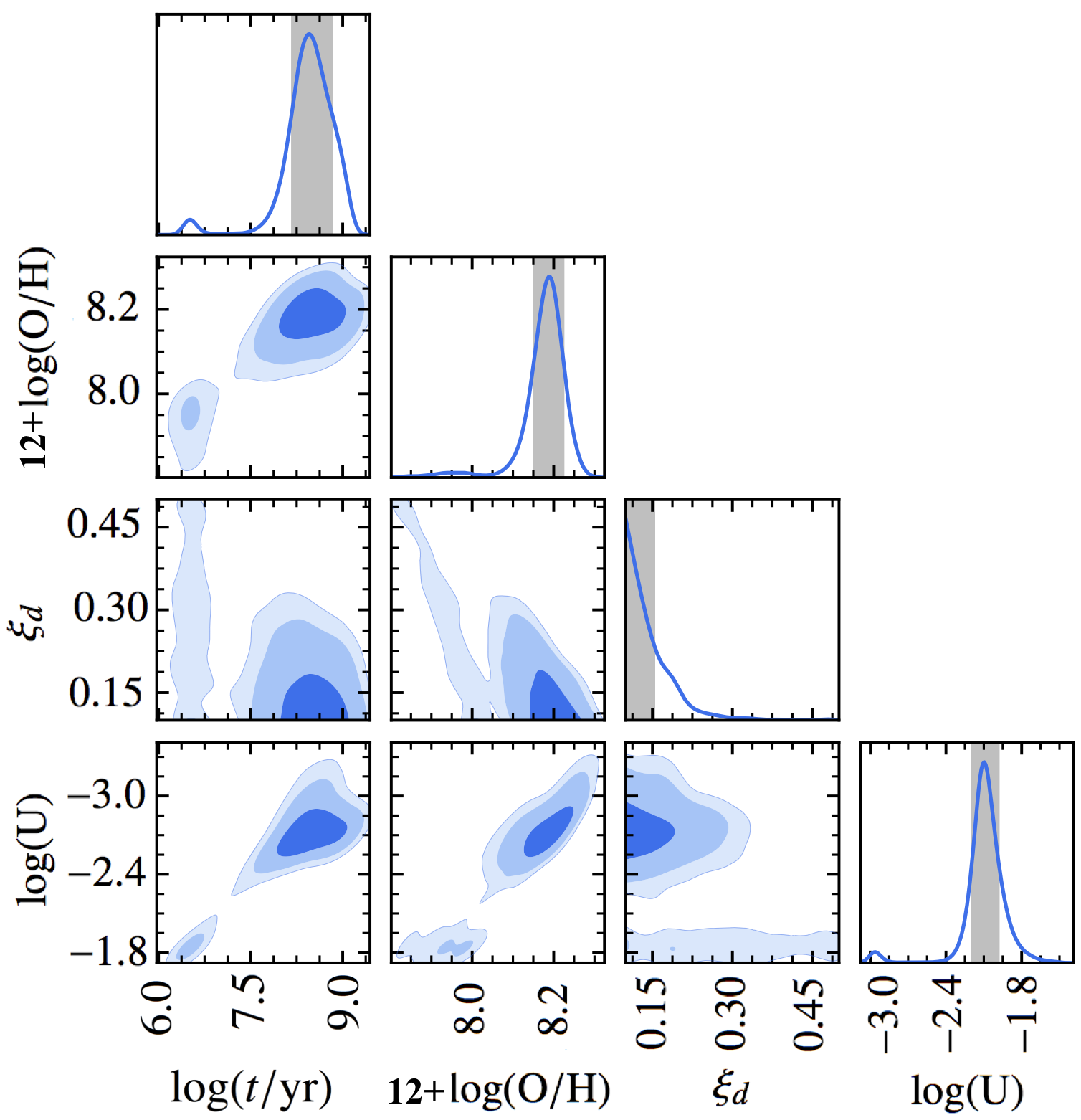}   
\caption{Triangle plot from fit to C\4~profile. Table~\ref{tab:priors} and the text which describes it explain what the properties which are plotted are. The shaded blue clouds show values which are one, two, and three $\sigma$ from the median. The shaded gray rectangles are values which are one $\sigma$ from the median.}
\end{figure}

In Figure~\ref{fig:test1_triangle}, we show the corresponding two-dimensional joint probability distribution functions (PDFs) and one dimensional marginal posterior PDFs of physical quantities. Hereafter, we refer to this type of plot as a triangle plot. The properties which are included in Figure~\ref{fig:test1_triangle} are the time since the beginning of star formation, the ionized-gas oxygen abundance, the dust-to-gas metal mass ratio, and the ionization parameter. We find that just fitting the stellar + nebular C\4~profile does not constrain any of the latter four quantities. In particular, the distributions are broad and/or double-peaked. When considering the median values of fitted parameters and comparing them to values of the physical properties which are reported in the literature, we find that the median-model time since the beginning of star formation, which is log(age/yr)$\sim7.8$ ($\sim63$\,Myr), is higher than the range of photometrically-derived ages for the individual SSCs within the COS aperture \citep{Reines2008, Adamo2010}; and that the median-model oxygen abundance, which is 12+log(O/H)$\sim8.2$, is significantly higher than values obtained via the standard method by \cite{Papaderos2006, Izotov2006} and  \cite{Izotov2009}.

\subsection{Tests 2 \& 3: fits to UV fluxes and equivalent widths}\label{sub:tests2and3}

As test 2, we try to simultaneously fit: i) the fluxes of all nebular high-ionization UV emission lines and ii) the equivalent widths (EWs) of the absorption and nebular emission components of  C\4\,$\lambda\lambda$1548, 1551. The wavelength limits which we use to obtain the flux and EWs are given in Table~\ref{tab:numerically_int_fluxes}. As test 3, we repeat test 2 but excluding the He\2\,$\lambda1640$ flux from the fit, as this is the highest ionization UV emission line which we detect. The results from tests 2 and 3 are summarized in Table~\ref{tab:uv_observables_vs_model}, which gives: line ID (column 1),  observed flux or EW (column 2), median-model flux or EW corresponding to test 2 (column 3), median-model flux or EW corresponding to test 3 (column 4), difference between columns 2 and 3 (column 5), and difference between columns 2 and 4 (columm 6). From test 2 we conclude that the residuals of all observables with the exception of the C\4~absorption EW,are within the observational uncertainty. From test 3 we conclude that excluding the He\2~flux from the fit does not significantly impact the results from test 2. In summary, we are unable to fit the C\4~absorption EW within the observational uncertainties. A main difficulty with trying to measure the C\4~EW is that it is weak and very broad. Thus, it is not clear where to put the wavelength limits for the numerical integration of this feature. 

Table~\ref{tab:uv_physical_properties} gives the median values of the physical parameters corresponding to tests 2 and 3. We do not expect our masses to be accurate since we are not fitting the stellar continuum. For reference, for the clusters which are within the COS aperture, \cite{Adamo2010} obtained a total stellar mass of $4.5\times10^6\,M_\odot$. For the time since the beginning of star formation we obtain 3.9$\pm0.4$ Myr via test 2 and 5.60$_{-1.6}^{+12.4}$ via test 3. For reference, \cite{Adamo2010} finds that the youngest cluster within our aperture is 6$\pm1$ Myr. For the ionized-gas oxygen abundance, we find that it is unrealistically-low, which is in agreement with the work presented in \cite{Kehrig2018}, where the authors fitted the flux of the integrated He\2~4686 emission-line of \sbs. Finally, for log(C/O), we find similar values via test 2 and 3, i.e. $\sim$-0.45 dex. This is higher than the value obtained by \cite{Garnett1995} via the standard method, which is log(C/O)=-0.94$\pm0.17$.

\begin{table}\label{tab:uv_observables_vs_model}

\centering
\caption{Observed versus median-model UV fluxes and equivalent widths. \emph{Columns.} (1) Line ID. (2) Either the logarithm of the observed flux (erg s$^{-1}$ cm$^{-2}$) corrected for foreground reddening only (top five rows), or the EWs (\AA) of the C\4~components which are described in subsection~\ref{sub:tests2and3} (last two rows). The values in this column are the same as the ones in Table~\ref{tab:numerically_int_fluxes}. (3) Either the median-model flux including intrinsic reddening, or the median-model EW, both for test 2. (4) Same as (3) but for test 3. (5) Residual between (2) and (3). (6) Residual between (2) and (4).}
\begin{tabular}{lccccc} 
\hline
Line ID   & Observed & Test 2 & Test 3 & $\Delta_1$ & $\Delta_2$ \\
(1)  &  (2)  & (3) & (4) & (5) & (6)  \\
\hline\hline
C\,{\sc iv}e 1550 & $-13.94\pm0.02$ & -13.95 & -13.95 & -0.01 & -0.01 \\
He\,{\sc ii} $\lambda$1640 & $-13.99\pm0.02$ & -14.00 & -  &  -0.01 & - \\
O\,{\sc iii}$]$ $\lambda$1661 & $-14.41\pm0.05$ & -14.45 & -14.46  & -0.04 & -0.05 \\
O\,{\sc iii}$]$ $\lambda$1666 & $-14.04\pm0.02$ & -14.02 & -14.03  & -0.02 & -0.01 \\
C\,{\sc iii}$]$ $\lambda$1909 & $-13.74\pm0.03$ & -13.72 & -13.72  & +0.02 & +0.02 \\
\hline
C\,{\sc iv}e 1550 & $1.70\pm0.20$ & 1.70 & 1.83  & 0.00 & 0.13 \\
C\,{\sc iv}a 1550 & $-0.90\pm0.10$ & -0.29 & -0.28  & 0.61 & 0.62 \\
\hline
\end{tabular}
\end{table}

\begin{table}\label{tab:uv_physical_properties}

\centering
\caption{Median physical properties derived from the UV data. For each property, we give the one-sigma errors around the median. \emph{Columns.} (1) Property. (2) Results from test 1. (3) Results from test 2.}
\begin{tabular}{lll} 
\hline
Line ID   & Test 2 & Test 3 \\
(1)  &  (2)  & (3)  \\
\hline\hline
    log(M$_\star$\,/\,$M_\odot$) & +6.24$_{-0.09}^{+0.13}$ & +6.31$_{-0.15}^{+0.40}$\\
    t / Myr & +3.90$_{-0.40}^{+0.40}$ & +5.60$_{-1.6}^{+12.4}$\\
    log(\Zism) & -3.99$_{-0.01}^{+0.01}$ & -3.95$_{-0.03}^{+0.07}$\\
    log(U$_{\rm S}$) & -2.01$_{-0.05}^{+0.06}$ & -1.90$_{-0.09}^{+0.16}$\\
    $\xi_{\rm d}$ & +0.47$_{-0.04}^{+0.03}$ & +0.44$_{-0.09}^{+0.04}$\\  
    12+log(O/H) & +6.45$_{-0.02}^{+0.03}$ & +6.52$_{-0.05}^{+0.08}$\\      
    $A_{1500}$ (mag) & +0.32$_{-0.22}^{+0.40}$ & +0.28$_{-0.20}^{+0.59}$\\ 
    $(\mathrm{C/O})/(\mathrm{C/O})_\odot$ & +0.79$_{-0.05}^{+0.05}$ & +0.81$_{-0.06}^{+0.07}$\\
\hline
\end{tabular}
\end{table}

\subsection{Tests 4 \& 5: fits to optical emission-line fluxes}\label{sub:tests4and5}

In the optical, we exclude He\2 $\lambda$4686 from all fits, as none of the models which we tried were able to reproduce the flux of this line. As test 4, we  try to simultaneously reproduce all of the optical lines which are shown in the first column of Table~\ref{tab:optical_observables_vs_model}. For the latter fit, columns (2) and (3) of Table~\ref{tab:optical_observables_vs_model} give the observed and median-model flux values. The residual between (2) and (3) is given in column (5). The latter column shows that neither the high-ionization emission lines of [Ar IV] $\lambda$4740, [S III] $\lambda$6310 and [S III] $\lambda$9068 nor the [N II] $\lambda\lambda$6548, 6584 lines are well reproduced. In fact, only the Balmer and He\1~lines are reproduced within the 2\% observational errors. As test 5 we try fitting only the fluxes of lines with fluxes stronger than or equal to that of the $[$S\2]~$\lambda$6717 line. The median-model fluxes corresponding to the latter fit are shown in column (4) of Table~\ref{tab:optical_observables_vs_model}. The residual between (2) and (4) is given in column (6). The result is very similar although the fits to $[$O\3]~$\lambda$4959 and $[$O\3]~$\lambda$5007 slightly improve.  


The median physical properties derived from simultaneous fits to optical emission-line fluxes are summarized in Table~\ref{tab:physical_properties_optical}. We do not expect the masses to be accurate since we are not fitting the stellar continuum. The ages are unrealistically young, which is required to produce hard ionizing photons and approach the observed fluxes of the high-ionization optical lines. The attenuation at 1500 \AA~is smaller than that derived from the UV. We do not expect the C/O value to be accurate since there are no carbon lines in the optical.

\begin{table}\label{tab:optical_observables_vs_model}

\centering
\caption{Observed versus median-model optical emission-line fluxes. \emph{Columns.} (1) Line ID. (2) Logarithm of the observed flux (erg s$^{-1}$ cm$^{-2}$) corrected for foreground reddening only. (3) Median-model flux corresponding to test 4. (4) Median-model flux corresponding to test 5. (5) Difference between columns 2 and 3. (6) Difference between columns 2 and 4.}
\begin{tabular}{lccccc} 
\hline
Line ID   & Observed   &   Test 4 & Test 5  & $\Delta_1$ & $\Delta_2$\\
(1) & (2) & (3) & (4) & (5) & (6) \\
\hline\hline
$[$Ar\4]~$\lambda$4740	&	-15.41	$\pm0.02$ &	-15.63	&	-15.54	&	0.22	&	0.13  \\
H$\beta$~$\lambda$4861	&	-13.38	$\pm0.01$ &	-13.38	&	-13.38	&	0.00	&	0.00  \\
$[$O\3]~$\lambda$4959	&	-13.33	$\pm0.01$ &	-13.39	&	-13.35	&	0.06	&	0.02  \\
$[$O\3]~$\lambda$5007	&	-12.86	$\pm0.01$ &	-12.91	&	-12.88	&	0.05	&	0.02  \\
He\1~$\lambda$5876	    &	-14.37	$\pm0.01$ &	-14.36	&	-14.36	&	-0.01	&	-0.01 \\
$[$O\1]~$\lambda$6300	&	-15.54	$\pm0.01$ &	-15.58	&	-	    &	0.04	&	-	  \\
$[$S\3]~$\lambda$6310	&	-15.56	$\pm0.01$ &	-15.40	&	-	    &	-0.16	&	-	  \\
$[$N\2]~$\lambda$6548	&	-15.93	$\pm0.02$ &	-16.22	&	-	    &	0.29	&	-	  \\
H$\alpha$~$\lambda$6563	&	-12.91	$\pm0.01$ &	-12.92	&	-12.91	&	0.01	&	0.00  \\
$[$N\2]~$\lambda$6584	&	-15.47	$\pm0.01$ &	-15.76	&	-	    &	0.29	&	-	  \\
$[$S\2]~$\lambda$6717	&	-15.08	$\pm0.01$ &	-15.12	&	-15.16	&	0.04	&	0.08	  \\
$[$S\2]~$\lambda$6731	&	-15.18	$\pm0.01$ &	-15.22	&	-15.26	&	0.04	&	0.08	  \\
$[$Ar\3]~$\lambda$7135	&	-15.09	$\pm0.01$ &	-15.14	&	-15.15	&	0.05	&	0.06  \\
$[$S\3]~$\lambda$9068	&	-14.84	$\pm0.01$ &	-14.70	&	-14.70	&	-0.14	&	-0.14 \\
\hline
\end{tabular}
\end{table}

\begin{table}\label{tab:physical_properties_optical}

\centering
\caption{Median physical properties derived from optical data. For each property, we give the one-sigma errors around the median. \emph{Columns.} (1) Property. (2) Results from test 4. (3) Results from test 5.}
\begin{tabular}{lll} 
\hline
Line ID   & Test 4 & Test 5 \\
(1)  &  (2)  & (3)  \\
\hline\hline
    log(M$_\star$\,/\,$M_\odot$) & +5.93$_{-0.01}^{+0.02}$ & +5.92$_{-0.01}^{+0.01}$\\
    t / Myr & +0.13$_{-0.03}^{+0.11}$ & +0.11$_{-0.00}^{+0.01}$\\
    log(\Zism) & -3.06$_{-0.01}^{+0.01}$ & -3.25$_{-0.01}^{+0.01}$\\
    log(U$_{\rm S}$) & -2.23$_{-0.01}^{+0.01}$ & -2.22$_{-0.01}^{+0.01}$\\
    $\xi_{\rm d}$ & +0.27$_{-0.03}^{+0.02}$ & +0.11$_{-0.00}^{+0.01}$\\  
    12+log(O/H) & +7.48$_{-0.01}^{+0.01}$ & +7.42$_{-0.01}^{+0.01}$\\      
    $A_{1500}$ (mag) & +0.05$_{-0.04}^{+0.08}$ & +0.01$_{-0.00}^{+0.02}$\\ 
    $(\mathrm{C/O})/(\mathrm{C/O})_\odot$ & +0.98$_{-0.02}^{+0.02}$ & +0.13$_{-0.02}^{+0.06}$\\
\hline
\end{tabular}
\end{table}

\section{Discussion}\label{sec:discussion}

\subsection{EWs of high-ionization UV emission lines: nearby versus distant systems}

\cite{Senchyna2017} present \hst~COS UV observations of 10 nearby star-forming galaxies with 12+log(O/H) between 7.81 and 8.48. They find that objects with 12+log(O/H) above 8.1 have: i) P-Cygni like C\4~$\lambda\lambda\,$1548, 1551 profiles with broad absorptions forming in the wind of O-type MS stars; and ii) broad He\2~$\lambda1640$ emission forming in the winds of WR stars. On the other hand, they find that objects with 12+log(O/H)$\le8.0$ have weak stellar C\4~$\lambda\lambda\,$1548, 1551  and He\2~$\lambda1640$ profiles and are dominated by nebular emission in [C\3], C\3]~$\lambda\lambda$1907, 1909, C\4~$\lambda\lambda\,$1548, 1551, and He\2~$\lambda1640$. They find seven objects with [C\3], C\3]~$\lambda\lambda$1907, 1909 emission, some of them reaching extremely high equivalent width values ($\sim10-15\,\AA$). Finally, they find that C\4~$\lambda\lambda\,$1548, 1551 and He\2~$\lambda1640$ have equivalent widths below 2\,\AA~and conclude that the detection of C\4~$\lambda\lambda\,$1548, 1551 at$>20\,$\AA~may require extremely metal poor stars and gas. In \sbs, which is more metal-poor that the previous sample, C\4~$\lambda\lambda\,$1548, 1551 shows a weak P-Cygni profile and nebular components. The EWs of the nebular [C\3], C\3]~$\lambda\lambda$1907, 1909 and C\4~doublet emissions are 5 and 1.7~\AA, respectively. The largest C\4~$\lambda\lambda\,$1548, 1551 EW which has been reported for a nearby starburst galaxy is $\sim10\,$\AA. This is for galaxy J104457, which has 12+log(O/H)=7.45 \citep{Berg2019b} and is at a very similar redshift (0.013) compared to \sbs.

\cite{Senchyna2019} present \hst~COS UV  observations of six star-forming XMPs with 12+log(O/H) between 7.4 and 7.7, i.e., lower than in the work of \cite{Senchyna2017}. They find stellar C\4~$\lambda\lambda\,$1548, 1551 in five of the galaxies and high [C\3]+C\3]~equivalent widths ($\sim11\,$\AA) in two objects. The highest EWs of C\4~$\lambda\lambda\,$1548, 1551 and He\2~which they find are 4.4 and 1.7 \AA, respectively. For \sbs~we find EWs of 1.7 \AA~for C\4~$\lambda\lambda\,$1548, 1551 and He\2~$\lambda$1640.  Although nebular C\4~$\lambda\lambda\,$1548, 1551 emission is present in some very high sSFR nearby systems, nearby systems do not reach the equivalent widths of C\4~which are observed in individual lensed systems at z>6. Undisputed detections of nebular emission in He\2~in predominantly star-forming systems at z>6 remain elusive \citep{Senchyna2019}. Objects located at z=2-4 with high values of EWs(C\3]), i.e., $\geq10$ \AA, are analysed in \cite{Nakajima2018}, who find that EWs(C\3])>20 \AA~can only be modelled with a combination of photoionization by stars and an AGN. 

\subsection{C/O}\label{sub:c2o} 

\cite{Berg2016} and \cite{Berg2019a} studied the C/O ratio derived via UV collisionally excited emission lines of a sample of 12 low metallicty galaxies with significant detections of carbon and oxygen lines. At low metallicity, (12+log(O/H)<8.0), they find that no clear trend is evident in C/O versus O/H. We find log(C/O)$\sim-0.45$ via tests 2 and 3, and 12+log(O/H) = 7.48 / 7.42 via tests 4 / 5, respectively. This is very similar to what \cite{Berg2016} report for galaxy J120122, which is log(C/O)$\sim-0.45$ and 12+log(O/H) = 7.45.

\subsection{JWST observation}

As a simple exercise, we placed \sbs\ at redshift $z=10$, and used the COS G130M + G160M + G185M spectrum to simulate an observation of this galaxy with the near infrared spectrograph (NIRSpec) onboard \textit{JWST}. For this purpose, the spectrum's flux was diluted using the luminosity distance for cosmological parameters $H_0=69.6$, $\Omega_M=0.286$, and $\Omega_{vac}=0.714$. In addition, we accounted for IGM absorption of UV photons using \cite{Madau1995} attenuation. We degraded the spectrum to match NIRSpec's prism resolution, i.e., $\Delta\lambda=15.5$~\AA~or $R\sim100$ at $1550\times(1+z)$~\AA. We used the exposure time calculator of the NIRSpec Microshutter Array (MSA) prototype. The latter assumes a point source centered in an open MSA shutter of 0.2"\,$\times$\,0.45", with background from two adjacent open shutters. Detecting the C\4~$\lambda\lambda\,$1548, 1551 doublet with a signal to noise ratio of 6 per resolution element would require an exposure time of $10^5$ s, similar to what is adopted for `deep' observations of the \textit{JWST}/NIRSpec Guaranteed Time Observations (GTO) program \citep{Chevallard2019}.

\section{Summary and conclusions}\label{sec:conclusion}

We present new \hst/COS UV spectroscopy and VLT/MUSE optical spectroscopy of one of the nearest most metal-poor starburst galaxies known, \sbs. The main results from the analysis of these and archival data follow.
\begin{enumerate}
    \item The C\4\,$\lambda\lambda$1548, 1551 doublet is composed of a P-Cygni like profile plus nebular components which are redshifted relative to the Al\2\,$\lambda1670$ line (subsection~\ref{sub:line_measurements}).     
    \item In spite of its high sSFR (log[$sSFR$/yr]=-8.13), and extremely low metallicty, we do not detect equivalent widths in excess of 10~\AA~for the high-ionization UV emission lines of He\2, O\3], [C\3],C\3], or C\4. Thus, the combination of the above two properties does not guarantee large equivalent widths of these UV lines (subsection~\ref{sub:line_measurements}).    
    \item We report the detection of blue-shifted O\5~$\lambda$1371 which could indicate the presence of VMSs in the galaxy (subsection~\ref{sub:cos}), as well as photospheric S\5\,$\lambda$1502 absorption at the 2$\sigma$ level. In addition, we report the tentative detection of broad N\3\,$\lambda4640$ and C\4/[Fe\3]\,$\lambda$4658 emission lines from WR stars and no detection of C\4 \,$\lambda$5808 (subsection~\ref{sub:wr}) in the vicinity of SSC3. Higher spatial resolution UV observations of \sbs~would help to better constrain the populations of WR and potential VMSs in the observed region.
    \item It is unlikely that the ULX which is reported in \citet[our subsection~\ref{sub:hmxb}]{Prestwich2013}, an IMBH (subsection~\ref{sub:imbh}), or an extremely low metallicity shock + precursor (subsection~\ref{sub:shocks}) are the explanation for the high-ionization UV and optical emission lines of \sbs. In addition, the SSP \texttt{BPASS v2.1} binary models with $M_{\rm up}=300$\,$M_\odot$ and nebular emission cannot reproduce the optical emission line ratios of~\sbs, while the similar single non-rotating models can.
    \item We use different sets of UV and optical observables to test constant star formation models with single non-rotating stars (we note that \texttt{BPASS v2.1} binary models with constant star formation and nebular emission are not publicly available). We use different sets of UV and optical observables to test constant star formation models with single non-rotating stars. The models include VMSs and account for the integrated light of stars, gas and dust. We could not combine the UV and optical observations for this purpose due to a flux mismatch between the UV and optical. Single non-rotating models which include VMSs are able to simultaneously reproduce all UV emission lines but only if the metallicity is unrealistically low (subsection~\ref{sub:tests2and3}). The same models are unable to reproduce the fluxes of the high-ionization optical emission lines. This is likely because in the models, the abundances of some of the corresponding elements might be wrong (subsection~\ref{sub:tests4and5}).
    \item We find that 12+log(O/H)$\,=7.45\pm0.04$ and log(C/O)$\,=-0.45^{+0.03}_{-0.04}$, (subsection~\ref{sub:c2o}) which is similar to the values of these abundance ratios for galaxy J120122 \citep{Berg2016}.
    \item Galaxies with the properties of \sbs~would be detected with \textit{JWST}/NIRSpec deep observations.
\end{enumerate}

\section*{Acknowledgements}

We acknowledge the referee for useful comments that improved the quality of this paper. AW acknowledges the support of UNAM via grant agreement PAPIIT no. IA105018. AW, AVG, AF, JC, and SC acknowledge the support of the ERC via an Advanced Grant under grant agreement no. 321323-NEOGAL. AVG acknowledges the support of the ERC via an Advanced Grant under grant agreement no. 742719-MIST. M. H. acknowledges the support of the Swedish Research Council, Vetenskapsr{\aa}det and the Swedish National Space Agency (SNSA), and is Fellow of the Knut and Alice Wallenberg Foundation. Based on observations collected at the European Southern Observatory under ESO programme 96.B-0690(A). We are extremely thankful to Christophe Morisset for his help with the shock + precursor models of \cite{Alarie2019}; to J. J. Eldridge and Elizabeth Stanway for their help with the BPASS models.

\section{Data availability}

The \hst~data underlying this article are available in the Mikulski Archive for Space Telescopes at \url{https://mast.stsci.edu/portal/Mashup/Clients/Mast/Portal.html}, and can be accessed with the dataset identifiers which are provided in column 1 of Table~\ref{tab:obs_log}. The reduced MUSE datacube used here and in \citep{Herenz2017b} is
available online via the CDS at http://vizier.unistra.fr/viz-bin/cat/J/A+A//606/L11.











\bsp	
\label{lastpage}
\end{document}